\documentclass{aa} 

\title{Taurus NESTs}
\usepackage{longtable} 
\usepackage{xcolor}
\usepackage{lscape}

\usepackage{graphicx}
\usepackage{txfonts}
\usepackage{hhline}

\usepackage{subfigure}
\usepackage{color}

\usepackage{natbib}

\bibpunct{(}{)}{;}{a}{}{,} 

\newcommand{\be}{\begin{equation}}
\newcommand{\ee}{\end{equation}}
\newcommand{\barr}{\begin{array}}
\newcommand{\earr}{\end{array}}

\begin{document}

    \title{Multiplicity and clustering in Taurus star forming region.
 }

 \subtitle{II. From ultra-wide pairs to dense NESTs}
  
\author{Isabelle Joncour\inst{1,2}\thanks{\emph{\tiny{Email:}} {\scriptsize isabelle.joncour@univ-grenoble-alpes.fr; joncouri@gaia.astro.umd.edu}}, Gaspard Duch\^ene\inst{1,3}, Estelle Moraux\inst{1} and,  Fr\'ed\'erique Motte\inst{1}  }

\authorrunning{Joncour et al.}

\institute{Univ. Grenoble Alpes, CNRS, IPAG, F-38000 Grenoble, France
  \and Department of Astronomy, University of Maryland, College Park, MD 20742, USA
  \and Astronomy Department, University of California, Berkeley, CA 94720-3411, USA} 
 
   \date{Accepted: August 31, 2018 }
\abstract
  {Multiplicity and clustering of young pre-main sequence stars appear as critical clues to understand and constrain the star formation process. Taurus is the archetypical example of the most quiescent star forming regions that may still retain primeval signatures of star formation.}
   {This work identifies local overdense stellar structures as a critical scale between wide pairs and loose groups in Taurus.}
   {Using  the  density-based spatial clustering of applications with noise ({\tt dbscan}) algorithm, and setting its free parameters based on the one-point correlation function and the k-nearest neighbor statistics, we have extracted reliably overdense structures 
from the sky-projected spatial distribution of stars.}
   {Nearly half of the entire stellar population in Taurus is found to be concentrated in 20 very dense, tiny and  prolate regions called NESTs (for Nested Elementary STructures). They are regularly spaced ($\approx 2$ pc) and mainly oriented along the principal gas filaments axes. Each NEST contains between four and 23 stars. Inside NESTs, the surface density of stars may be as high as 2500 pc$^{-2}$ and the mean value is 340 pc$^{-2}$.  
 Nearly half (11)  of  these NESTs contain about 75\% of the class 0 and I objects. The balance between  Class I, II, and, III fraction  within the NESTs suggests that they may be ordered as an evolutionary temporal scheme, some of them getting infertile with time, while other still giving birth to young stars.   
 We have inferred that only 20\% of stars in Taurus do not belong to any kind of  stellar groups (either multiple system, ultra wide pairs or NESTs). 
  The mass-size relation for stellar NESTs is very close to the Bonnor-Ebert expectation. The range in mass is about the same as  that  
  of dense molecular cores. The distribution in size is bimodal peaking at  12.5 and 50 kAU and the distribution of the number of YSOs in NESTs as a function of size exhibits two regimes.}
{We propose that the NESTs in their two size regimes represent the spatial imprints of stellar distribution at birth as they may have emerged within few millions years from their natal cloud either from a single core or from a chain of cores. We have identified them as the preferred sites of star formation in Taurus. These NESTs are the regions of highest stellar density and intermediate spatial scale structures between ultra-wide pairs and loose groups.   }

\keywords{Methods: statistical, data analysis -- binaries: visual -- Stars: formation -- Stars: pre-main sequence -- Stars: statistics -- Galaxy: open clusters and associations: individual: Taurus}

   \maketitle

\section{Introduction}
The Taurus star-forming region is the archetype of a quiescent and sparse star distribution associated to a low surface stellar density ($\rho \sim 0.8 \, {\rm pc}^{-2}$,  $\sim 5 \,\deg^{-2}$), spread over an area of $\approx\,$420\,pc$^2$ on the sky for the central part  in Taurus. A high-multiplicity fraction and loose spatial clustering are key features in this region. On the clustering side, it is now well established that
the stellar population of Taurus is distributed in a few  groups of 1.5-2.5 pc size  
\citep{JonesHerbig1979,GomezEtAl1993,KirkMyers2011}. On the multiplicity side, the companion frequency of young stars in Taurus is generally twice that of field stars \citep{DucheneKraus2013}. Moreover the presence of most probably coeval ultra-wide binaries appears to be a crucial outcome of the star formation process in that region \citep[hereafter Paper\,I]{JoncourEtAl2017}, a conclusion that also applies to the Perseus region for the embedded young stellar objects (YSOs) \citep{SadavoyStahler2017}. The higher order multiplicity property  of these ultra-wide pairs (UWPs) in Taurus suggests a close link  between the spatial scales associated  with multiple systems and  with UWPs, $\lesssim$1\,kAU and $\lesssim$60\,kAU respectively. Indeed, the UWPs population appears to extend the traditional multiple system regime by almost two orders in magnitude in spatial scale.
In Paper\,I, we have proposed that this property  illustrates the remaining imprints of a molecular core or clump fragmentation cascade scenario that  gave birth to young stars. 
Many of the UWPs in Taurus are made of two multiple systems, and are therefore "small stellar groups" of up to five members. However, the method used to identify UWPs was based on the mutual nearest neighbor property, which could only identify pairs. Higher order multiplicity (groupings of three or more separate systems) were thus not addressed in Paper\,I. We now wish to detect local stellar groups (overdensities) of any order, and thus need to use a different methodology. We have adopted a clustering algorithm, which is appropriate for this task. 

In  continuity with  
Paper\,I, the present study aims at analyzing spatial structures within the Taurus stellar population at intermediate length scales. In order to reach this objective, we propose a new methodology based on bottom-up approach using local physical properties to uncover explicitly spatial structures at this scale with a high level of confidence. The paper is organized as follow: in Section\,\ref{sec:methods}, we describe the methodology that allows the identification  with high reliability of the new spatial structures that we called NESTs (Nested Elementary STructures) described in the Section\,\ref{sec:nests}. In Section\,\ref{sec:props}, we characterize those structures and highlight their fundamental properties, while in Section\,\ref{sec:discus}, we discuss further properties and the nature of the NESTs and give our conclusions in Section\,\ref{sec:conclu}.

\section{Method}\label{sec:methods}

\subsection{Data}
 
We used the same Taurus catalog  as described in Paper\,I. It contains 338 members of Taurus taken from the full census of members down to $0.02\,M_{\odot}$ \citep{LuhmanEtAl2010} that we complement by the multiplicity information, including at high angular resolution wherever the information was available with a high degree (67\%) of  completeness.  Following the seminal work of \cite{Larson1995}, we aim at distinguishing clustering from multiplicity. To this end, we grouped together all stars that are separated by less than 1 kAU to form a single entity and  simply called them “multiple systems”. While this threshold is somewhat arbitrary, we selected it in paper\,I based on two complementary arguments. First of all, it is the lower threshold that defines by consensus wide binaries with separation larger than 1 kAU  \citep{ReipurthMikkola2012,Tokovinin2017}). Second, this threshold is close to the beam (seven arcsec) of the Spitzer-MIPS instrument, which is used in the classification of young stellar objects, particularly  the most embedded ones. Therefore, the census of neighbor stars is completely independent of the Class of the Taurus members for all separations larger than this threshold.

We note that \citet{KrausEtAl2017} have presented a more recent catalog that includes a new distributed population of diskless stars. The newly identified members are distributed over a broader footprint and belong primarily to the dispersed stellar population, which is typically older than the "classical" stellar population of Taurus. Therefore leaving this population aside does not significantly affect our conclusions surrounding local overdense stellar regions and we retain the catalog from Paper\,I for our analysis. 

\subsection{Main approaches to identify subgroups in SFR}

The typical  tools that are used to study the stellar spatial distribution may be subdivided in two main categories. The first focuses on spatial analyses in a global sense, aiming at characterizing the sources distribution as a whole (e.g., the  two-point correlation function is used to evaluate the degree and the regimes of clustering,  and the one-point correlation function to probe the binary regime range). The second approach aims at extracting (sub-)structures as topological entities using clustering algorithms, to further characterize them and derive their geometrical and physical properties. This task of finding clusters in a set of points has a long history in the field of applied mathematics and computer science. But intuitively, clusters can be seen simply as regions of enhanced stellar density with respect to their surroundings. Despite this simple definition, decades of dedicated and still on-going work show that there exist several methods to identify clusters (see appendix \ref{AppSec:ChoiceClustAlgo} for a short review and references therein). Each of  these methods has its pros and cons, and the optimal algorithm depends heavily on the type of data at hand and on the scientific goals.

The global distribution analysis and the substructures identification are complementary studies and, in some cases, the same tool can be used in both approaches. In particular, this is the case of the Minimum Spanning Tree \citep[MST,][]{Kruskal1956,Prim1957, Dijkstra1960,Zahn1971}. This method is a graph procedure that connects all points in an ensemble through a simple path such that the total distance length of its edges is minimal. The method then generates clusters by deleting the longest individual segments of the MST above a heuristically chosen threshold, such that clusters are defined as the remaining connected subgraphs. 
In the field of astronomy, the MST method was first applied in extragalactic astrophysics by \cite{BarrowEtAl1985}. 
Over the last decade, it is widely used in the context of young star clusters to identify subgroups of young stars, although 
alternate probabilistic techniques have been proposed recently \citep[e.g.,][]{KuhnEtAl2014}.

\cite{GutermuthEtAl2009} have proposed an empirical procedure to identify a critical MST length threshold above which the MST edges may be removed to identify clusters (plus usually a user-based condition on the minimum objects that must be located in the subgroups to be considered as such). This technique was notably used by 
\cite{KirkMyers2011} to study the stellar groups properties in four star-forming regions (Taurus, Lupus\,III, Cha\,I, and IC\,348). 

The MST method can also be used to define useful parameters to quantify the degree of sub-structuring, such as the $Q$ ratio of the average branches length of the MST over the mean separation of all pairs of stars \citep{CartwrightWhitworth2004,SchmejaEtAl2008,WrightEtAl2014,parker2014}, or to investigate mass segregation \citep{AllisonEtAl2009}. 
A comparison and discussion on different methods assessing a global measure of the sub-structuring degree and the mass segregation can be found in \cite{schmeja2011}, \cite{KupperEtAl2011}, and \cite{ParkerGoodwin2015}. 
The strength of the MST method is that it is a non-parametric technique, but its major drawbacks are inherent to all single linkage techniques, namely the artificial chaining effect, and the impossibility of handling noise and outliers.

\subsection{dbscan}  

In this work, we have chosen to identify spatial features at a given scale using local density properties and a connectivity rule to link together adjacent stars having similar stellar density neighborhood. 
The spatial structures thus identified are defined as connected "density cluster", following previous graph-principled work \citep{Hartigan1975}. To efficiently identify density clusters, we have chosen the non-parametric, one-level {\tt dbscan} (Density-Based Spatial Clustering Applications with Noise) clustering algorithm  developed in the Knowledge Discovery Database field \citep[see appendix \ref{AppSec:Dbscan}]{EsterEtAl1996}.
{\tt dbscan} has several advantages over other clustering algorithms (see appendix \ref{AppSec:ChoiceClustAlgo}): the partition of stars within density components is unique, it allows the  detection of clusters of arbitrary shape and size at a global scale from local requirements, and it is the only clustering algorithm that  explicitly labels noise and outliers. While other clustering algorithms induce a complete partitioning of the ensemble, {\tt dbscan} proposes partial clustering. Thus, unlike single linkage and MST algorithms, it is resistant to noise and outliers. The key algorithmic idea of {\tt dbscan} is to incrementally group stars, provided that (1) they are direct or indirect neighbors, and (2) the stellar density neighborhood of neighbors satisfied the selected criteria. {\tt dbscan} directly searches for connected dense regions in space separated by local stellar density drops. 

In {\tt dbscan}, the "density cluster" $C_{\epsilon,N_{min}}$ sets are determined by the choice of two local free parameters: a distance $\epsilon$ and a number of stars $N_{min}$. Stars are grouped together if they satisfy two conditions: (1) all stars within a set $C_{\epsilon,N_{min}}$ have a minimum number of stars $N_{min}$ within a radius $\epsilon$, (2) each star within a set $C_{\epsilon,N_{min}}$ is connected to any other star of the same set by a sequence of neighbors separated at most by a distance $\epsilon$.  
With these properties, a density cluster $C_{\epsilon,N_{min}}$ is said to be maximal among connected sets, i.e., $C_{\epsilon,N_{min}}$ is not contained in any larger density cluster defined at the same local $\epsilon$ level. 
As we explain next, the values of $\epsilon$ and $N_{min}$ that we use in this study have been determined based on the confidence level that local structures are distinct from random fluctuations.

\subsection{Selecting the algorithm parameters}

The clustering property in Taurus is traditionally  characterized using the two-point correlation function \citep[TPCF,][]{GomezEtAl1993,Larson1995,Simon1997,GladwinEtAl1999,Hartmann2002,KrausHillenbrand2008}. This method reveals an "elbow" in the 4--40\,kAU range, which is interpreted either as the signature of the Jeans instability in cool dense molecular cores \citep{Larson1995} or as an indication for a quasi-constant surface density of pairs produced by random motions that smooth out  primordial stellar lumps \citep{KrausHillenbrand2008}. Either way, the TPCF does not yield a unique plausible value for $\epsilon$ and we thus have sought another approach.

In Paper\,I, we have introduced the one point correlation function  defined as the probability of having a closest star located at $r$ from any chosen star at random. It provides a local analysis over all the range of $r$. This function differs and complements the  two-point correlation function, the latter being the probability of having pairs separated by $r$. At small $r$, these two functions describe a same spatial property. At larger $r$, the  two-point correlation function gives an "integrated" view of the spatial characterization of the spatial distribution. We defined 
\be
\Psi(r) = \frac{w_T(r)}{w_R(r)}, 
\ee
as an estimator of the one-point correlation function, where $w_T$ is the distribution of nearest neighbor distances in Taurus, and $w_R$ the same distribution for a random distribution with the same mean surface density.

\begin{figure}[!ht]
\includegraphics[width=\columnwidth]{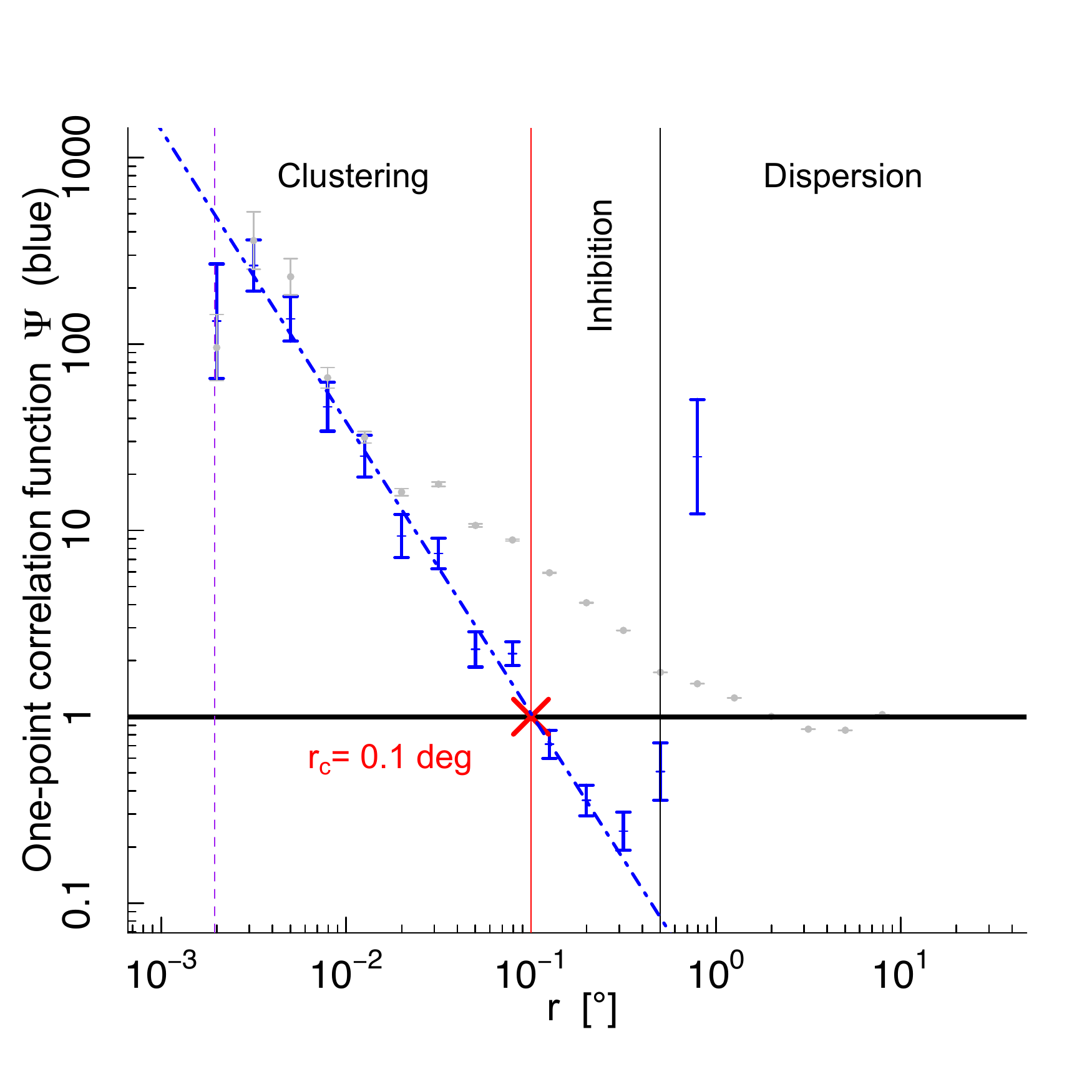}
\caption{One-point correlation function $\Psi$ (blue symbols and blue dot-dashed line fit) and estimated
pair correlation function g (gray symbols). This figure is reproduced from Paper\,I in order to highlight the choice of the $\epsilon$ parameter as $r_c$, the intersection point  between the horizontal line (random expectation)  and the $\Psi$ function. This value indicates the scale at which we want to investigate local overdense structures.
\label{Fig:OneTwoPoint_rc}}
\end{figure}

This $\Psi$ function, which encapsulates the coarsest trends 
in the stellar spatial distribution of star, reveals 
three different spatial regimes in Taurus (clustering, inhibition and dispersion). 
The clustering regime extends over all distances associated with an excess of stars that have a nearest neighbor  less than $r_c =0.1^\circ\approx 0.24$\,pc ($\approx 50$\,kAU) over a random distribution,  as seen in  Figure\,\ref{Fig:OneTwoPoint_rc} (adapted from Figure 4 in paper\,I). This introduces a natural benchmark for the local scale around which the value of $\epsilon$ has to be set.

We have endeavored to set the two free parameters $\epsilon$ and $N_{min}$ of the {\tt dbscan} algorithm by requiring that local overdense features are detected at a high level of significance ($\alpha = $99.85\%, i.e., three $\sigma$, if the distribution was to be normal) above random expectation. In Paper\,I, we have derived  the theoretical cumulative distribution $\mathcal{W}_k(r)$  of the $k$-nearest neighbor distribution in the case of a 2D random distribution:
\begin{equation}
\barr{lll}
\mathcal{W}_k(r)  &= &   2 (\pi \rho)^k/\Gamma(k) \cdot r^ {2k-1}\cdot \mathrm{exp}(-\rho \pi r^2)  \\
\mathcal{W}_1(r)  &= & 1 -\exp(-\pi \rho r^2) \\
\mathcal{W}_2(r)  &= & 1- (\pi \rho r^2+1)\, exp(-\pi r^2)\\
\mathcal{W}_3(r)  &= & 1/2 \left[ exp(-\pi r^2) \left( -\pi \rho r^2(\pi \rho r^2 +2)-2 \right)+2 \right], \\
\earr
\label{Eq:NND_k1_cdf_theo}
\end{equation}
where $\Gamma$ is the Gamma function, $\rho=5$ stars/${\rm deg}^2$ is the mean stellar surface density of Taurus, and $r$ is the distance to the $k$-nearest neighbor.
The value of $\mathcal{W}_k(r)$ as described in eq. \ref{Eq:NND_k1_cdf_theo}, gives the probability for a random star to have a k-nearest neighbor located at a distance of $r$ or less. In turn, $\alpha=1-\mathcal{W}_k(r)$ represents the degree of significance of an over-density defined by k+1 stars. 
Setting $r_c=0.1^\circ$ as the relevant local length scale, we see  in Figure\,2 that the probability of having a first companion (i.e., two stars within $r_c$) is fairly high, i.e., $\mathcal{W}_1(r_c)\approx 0.15$. Similarly, the probability of having 2 companions within $r_c$ is $\approx 0.11$. We must therefore consider the third nearest neighbor cumulative distribution. With $N_{min}=3+1=4$, we must set $\epsilon = 0.112^\circ$, very close to the previously identified length scale benchmark $r_c$, to achieve the required confidence level.
\begin{figure}[!ht]
\includegraphics[width=\columnwidth]{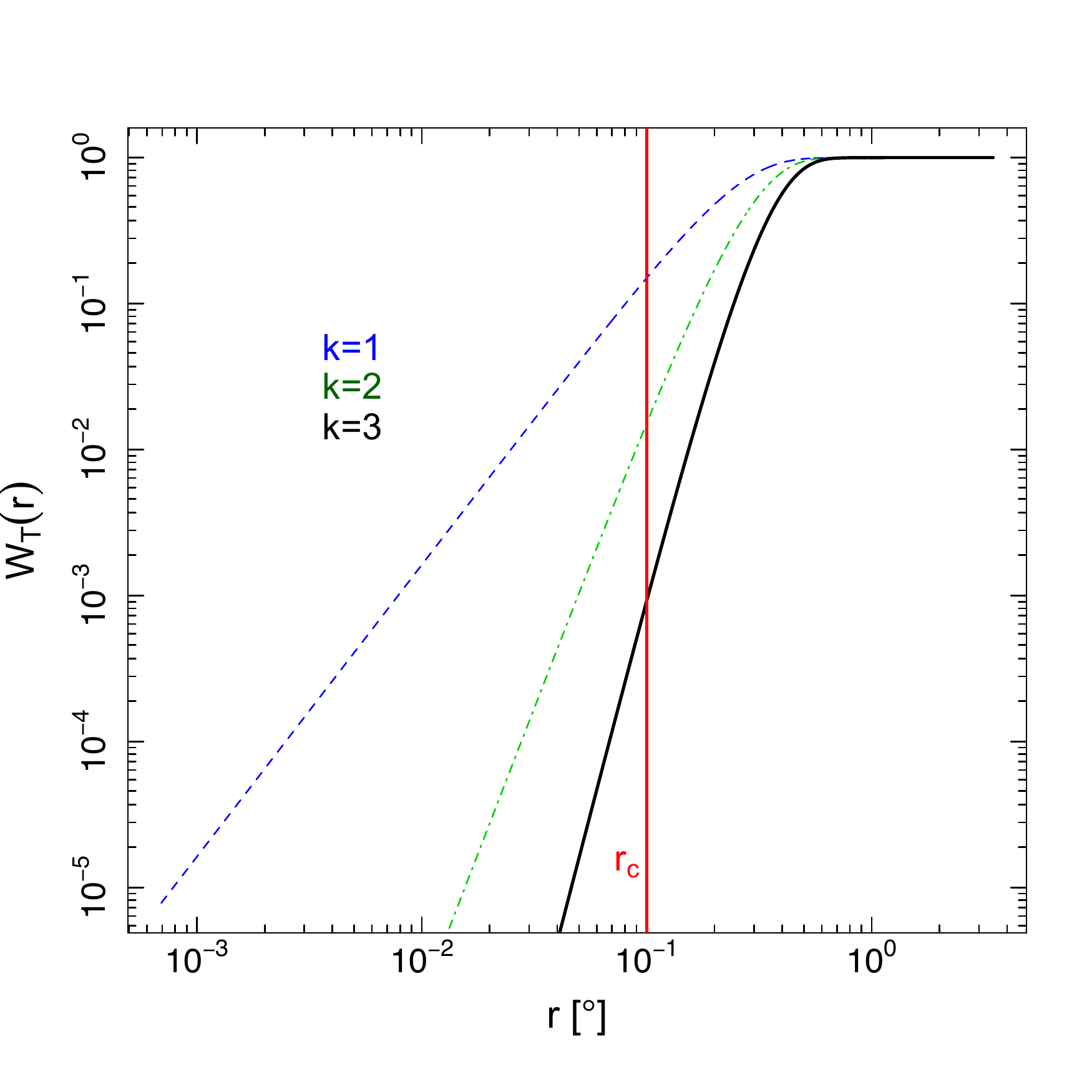}
\caption{Cumulative distribution of the first nearest neighbor ($W_1(r)$, dashed blue), second nearest neighbor ($W_2(r)$, dot-dashed green) and the third nearest neighbor ($W_3(r)$, solid black) in a random spatial distribution. The solid red vertical line  is the $r_c$ value.
\label{Fig:ChoiceParam}}
\end{figure}
In summary, the two free parameters of the {\tt dbscan} agorithm have been set based on the nearest neighbor statistics analysis, with a requirement to identify local spatial features with a 99.85\% level of significance above random expectations. 

\section{NESTs detection}\label{sec:nests}
\subsection{A new type of structures in Taurus}
\label{sub:stelNESTs}

\begin{figure*}[!ht]
\includegraphics[width=\textwidth]{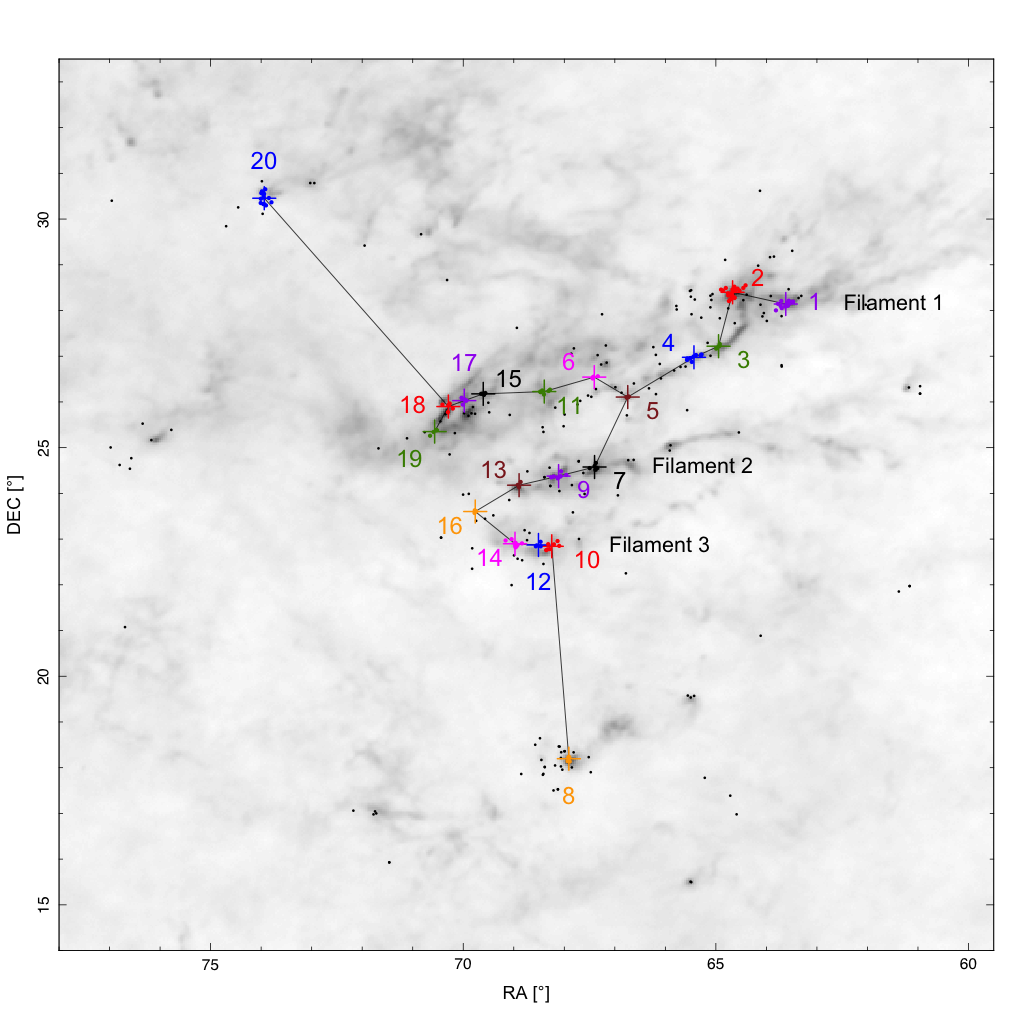}
\caption{Spatial distribution of the 20 NESTs identified in this study. The barycenter of each NEST is indicated by a colored plus mark and all the members are colored the same way. Star not pertaining to any NEST are shown as black dots. The MST defined by the NEST is shown with solid gray lines. NEST are numbered based on their increasing right ascension order (see Table\,\ref{Tab:NESTsCaract}). The Planck dust emission (217 Ghz) is mapped as the background.\label{Fig:TauG_NESTs_20_MST}}
\end{figure*}

With the parameters $\epsilon$ and $N_{min}$ set to the values defined in the previous section, we ran {\tt dbscan} on the Taurus catalog. We have identified 20 distinct stellar overdense structures, which we have termed Nested Elementary STructures (NESTs). Each NEST shelters between four stars -- the minimum number of stars imposed by the method -- and 23 stars, with a mean (resp. median) number of eight (resp. six) stars (see Table\,\ref{Tab:NESTsCaract}, Figure\,\ref{Fig:TauG_NESTs_20_MST} and Appendix\,\ref{AppSec:StarsInNests}). Considered as an ensemble, the NESTs contain nearly half of the entire stellar population in Taurus. Eighteen of these NESTs are located in the three principal gas filaments of the Taurus molecular cloud. 

To estimate the size of each NEST, we have first defined its convex hull as the smallest convex set of points that contains all of its members. The polygonal window drawn from this set of points provides an estimate of the minimal area $A$ enclosing all systems within the NEST. We have then evaluated the typical radius of each NEST as $R_{H}=(A/\pi)^{1/2}$; they range from $\approx5$ to $\approx80$\,kAU. The average of the first nearest neighbor separation (1-NNS) within all the NESTs, $r_{1}$, is about $20$\,kAU (i.e., 0.1\,pc). Thus NESTs appear as a new, intermediate, type of structure, on similar physical scales to UWPs but containing more elements, yet significantly smaller and denser than the already identified loose groups.

\begin{table*}[!ht]
\centering
{
\small
\begin{tabular}{r|rrrr|c|ccccc|ccc|rr}
  \hline
 N  & $n_*$ & $m_*$ & $\alpha_B$ & $\delta_B$ & $\bar{\rho}$ & 
$R_H$ & $a$  & $a/b$ & $e$ & PA & $r_N$ & $\alpha_N$ & $\beta_N$ & Fil. & $G_{KM}$\\ 
 & & [$M_\odot$] & [\degr] & [\degr] & [pc$^{-2}$] & [kAU] & [kAU] &  & & [\degr] & [kAU] & & & & \\
\hline
1 & 15 & 8.24 & 63.62 & 28.14 & 70 & 53.8 & 112.7 & 2.60 & 0.92 & 119 & 18.2 & 19.2 & 3.5 & 1 & 1 \\ 
  2 & 23 & 13.29 & 64.67 & 28.40 & 48 & 81.0 & 155.4 & 1.82 & 0.84 & 117 & 17.5 & 20.0 & 4.4 & 1 & 2 \\ 
  3 &  4 & 1.84 & 64.95 & 27.22 & 533 & 10.1 & 37.4 & 5.67 & 0.98 & 143 & 18.4 & 7.9 & 5.1 & 1 & 3 \\ 
  4 & 10 & 4.69 & 65.43 & 26.98 & 83 & 40.5 & 91.3 & 2.92 & 0.94 & 115 & 15.9 & 9.1 & 4.6 & 1 & 3 \\ 
  5 &  5 & 4.83 & 66.74 & 26.11 & 2 453 & 5.3 & 15.3 & 3.52 & 0.96 & 111 & 5.0 & 69.5 & 13.3 & 1 & - \\ 
  6 &  4 & 1.65 & 67.41 & 26.54 & 437 & 11.1 & 34.6 & 4.14 & 0.97 & 111 & 15.5 & 22.2 & 7.8 & 1 & - \\ 
  7 &  5 & 2.73 & 67.40 & 24.58 & 75 & 29.9 & 49.1 & 1.40 & 0.70 & 133 & 33.3 & 6.8 & 2.4 & 2 & - \\ 
  8 &  6 & 7.73 & 67.91 & 18.20 & 360 & 15.0 & 31.7 & 2.46 & 0.91 & 179 & 10.8 & 211.6 & 6.2 & - & 4 \\ 
  9 &  8 & 4.11 & 68.11 & 24.38 & 65 & 40.9 & 87.0 & 2.64 & 0.93 & 108 & 21.2 & 10.6 & 2.9 & 2 & 5 \\ 
  10 &  7 & 3.67 & 68.25 & 22.84 & 45 & 46.1 & 89.0 & 2.35 & 0.91 & 115 & 36.4 & 1.8 & 4.2 & 3 & 6 \\ 
  11 &  6 & 3.85 & 68.40 & 26.23 & 105 & 27.8 & 59.6 & 2.40 & 0.91 & 110 & 18.8 & 21.3 & 9.4 & 1 & - \\ 
  12 &  4 & 1.61 & 68.51 & 22.87 & 137 & 19.9 & 39.8 & 1.69 & 0.81 & 137 & 28.6 & 2.3 & 3.7 & 3 & 6 \\ 
  13 &  4 & 2.07 & 68.89 & 24.18 & 324 & 12.9 & 34.4 & 2.93 & 0.94 & 159 & 17.2 & 16.5 & 7.7 & 2 & - \\ 
  14 & 13 & 9.81 & 68.98 & 22.90 & 82 & 46.4 & 92.7 & 2.37 & 0.91 & 109 & 18.3 & 7.2 & 3.2 & 3 & 6 \\ 
  15 &  4 & 1.46 & 69.60 & 26.18 & 447 & 11.0 & 25.5 & 2.21 & 0.89 & 97 & 17.1 & 8.8 & 5.5 & 1 & - \\ 
  16 &  4 & 0.66 & 69.76 & 23.60 & 795 & 8.3 & 17.5 & 1.85 & 0.84 & 165 & 12.0 & 34.7 & 7.3 & 2 & - \\ 
  17 &  5 & 2.07 & 69.98 & 26.03 & 131 & 22.8 & 44.0 & 1.55 & 0.76 & 170 & 22.3 & 4.5 & 4.1 & 1 & 7 \\ 
  18 &  6 & 2.56 & 70.29 & 25.90 & 48 & 41.1 & 61.6 & 1.36 & 0.68 & 140 & 28.8 & 3.5 & 1.1 & 1 & 7 \\ 
  19 &  5 & 3.00 & 70.57 & 25.35 & 515 & 11.5 & 53.1 & 11.14 & 1.00 & 133 & 18.4 & 12.1 & 3.4 & 1 & 7 \\ 
  20 & 13 & 11.81 & 73.94 & 30.46 & 47 & 61.5 & 98.1 & 1.58 & 0.77 & 176 & 24.9 & 108.1 & 3.7 & - & 8 \\ 
\hline
\end{tabular}
}
 \caption{Properties of the NESTs. For each NEST, we list the number of stars it contains ($n_*$), its total stellar mass ($m_*$), the equatorial coordinates of its center ($\alpha_B$, $\delta_B$), its average stellar density ($\bar{\rho}$), its mean radius as computed from the area of the associated convex hull ($R_H$), the semi-major axis, aspect ratio, eccentricity and position angle of the associated ellipsoid hull ($a$, $a/b$, $e$ and PA), the mean 1-NNS distance inside the NEST ($r_N$), the separability and dilution factors ($\alpha_N$ and $\beta_N$), the filament it is associated with (Fil) and the loose group number it belongs to as defined in \citet[][$G_{KM}$]{KirkMyers2011}.
\label{Tab:NESTsCaract}}
\end{table*}

\begin{figure}[!ht]
\includegraphics[width=\columnwidth]{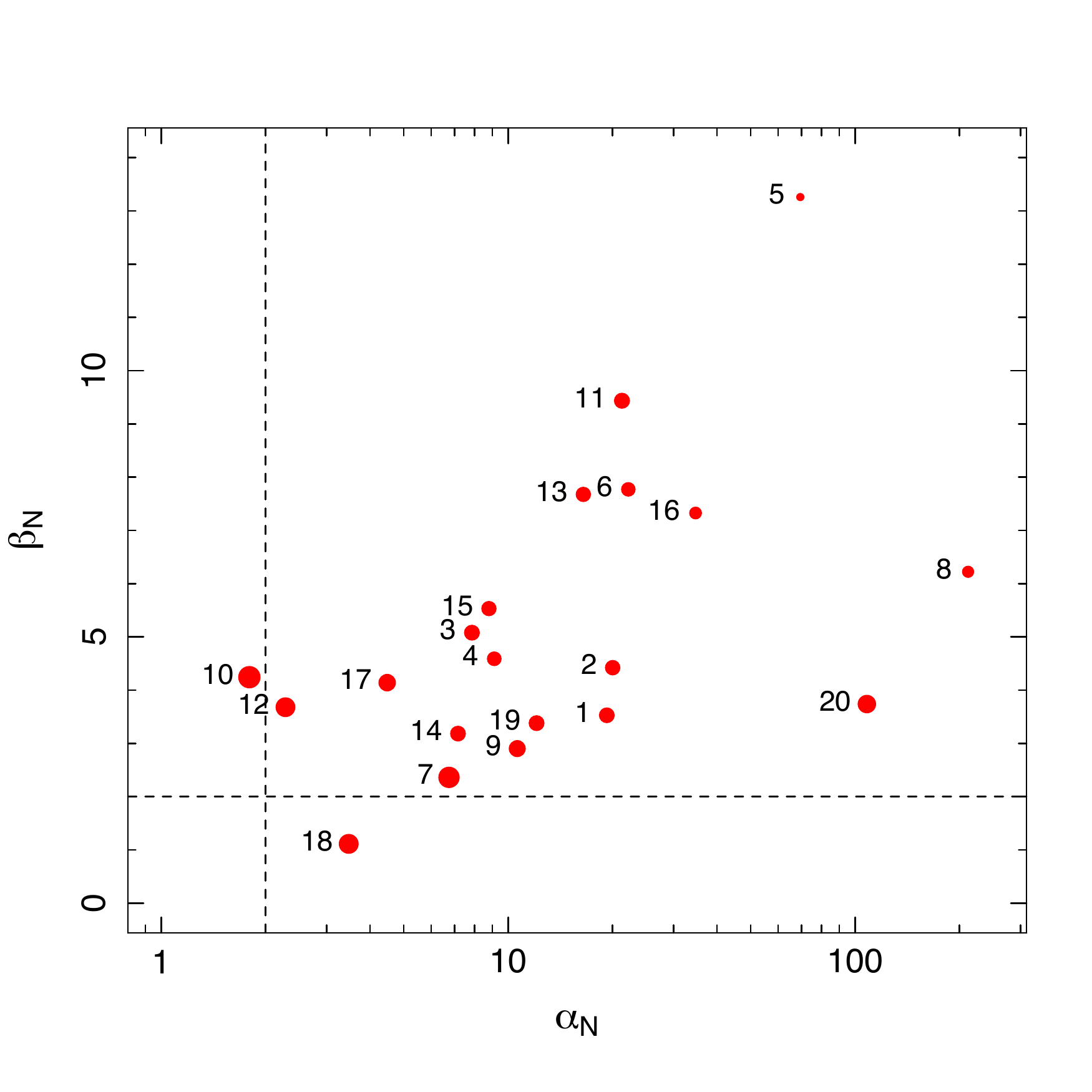}
\caption{Dilution factor $\beta_{N}$ as a function of the inter-NEST separation ratio $\alpha_{N}$. The size of the marks is indicative of the mean first nearest neighbor $r_N$. These ratios are markers of the reliability of the NESTs. Values higher than 2  (dashed black lines) indicate that a NEST is well separated both from the other NESTs and from the  stellar population outside the NESTs.
\label{Fig:TauG_NEST_sepTest_alpha_beta}}
\end{figure}

\subsection{Reliability of the NESTs}

Algorithms designed to identify overdense substructures within an ensemble of objects can sometimes produce spurious groupings and incorrectly split large groups into smaller sub-units. To test the reliability of these substructures, we have first verified their separability (i.e., how distinct NESTs really are from one another) and their degree of dilution (i.e., how significantly does the overdensity stands out relative to the  stellar population  outside the NESTs).

To test the separability between NESTs, we have compared the internal spacing of stars belonging to a single NEST to their distance to any other NESTs. The evaluation of this separability for each NEST is performed  by introducing a separability ratio $\alpha_{N}=D_{N}/r_{N}$, defined as the ratio of the nearest distance ($D_{N}$) from a star in a given NEST to a star belonging to any other NEST over the mean 1-NNS within the NEST ($r_{N}$). Large values of the ratio indicate well-separated NESTs. All but 2 NESTs have $\alpha_N \gtrsim 3$, with a median value of $\approx 11$. In other words, stars in a NEST are on average one order of magnitude closer to other members of the same NEST than to stars in other NESTs. NESTs number  10 and 12 are the least separated NESTs, raising the possibility that they are two substructures within a larger one.

In a second test, we have evaluated the degree of dilution of the NESTs relative to the more dispersed population, that is the stellar population outside the NESTs. Specifically, for a given NEST, we have first computed the nearest distance $d_*$ between a NEST member and a star of the dispersed population and we have defined the dilution parameter $\beta_N=d_*/r_N$. 
Smaller values of $\beta_N$ indicate that it is harder to distinguish a NEST from the surrounding population, as the overdensity of the NEST becomes increasingly marginal. Values for $\beta_N$ range from $\approx 1$ to $\approx 13$, with a median of 4. Therefore, NESTs are not in complete isolation but rather immersed in the more dispersed population. 

As Figure\,\ref{Fig:TauG_NEST_sepTest_alpha_beta} shows, the $\alpha_N$ and $\beta_N$ quantities are positively correlated ($p$-value of $10^{-2}$ based on nonparametric Spearman rank correlation test). Looking at the symbol size that is proportional to the mean 1-NNS, we note that the quantities $r_N$ and $\alpha_N$ on the one hand, and $r_N$ and $\beta_N$ on the other hand, appear significantly anti-correlated ($p$-value of $10^{-3}$ for both quantities). 
This indicates that the more compact a NEST is, the further it is from the nearest NEST and the more its members are separated from the dispersed population. Taking these correlations together, we conclude that there is a connection between the internal local density of the NEST and the local density of the immediate stellar neighborhood. Consequently, the NESTs and the dispersed population are somehow connected and must be interpreted in a comprehensive model.

\subsection{Robustness of the NESTs.}
\label{Sub_sec:Robust_NESTs}
\begin{figure}[!ht]
\includegraphics[width=\columnwidth]{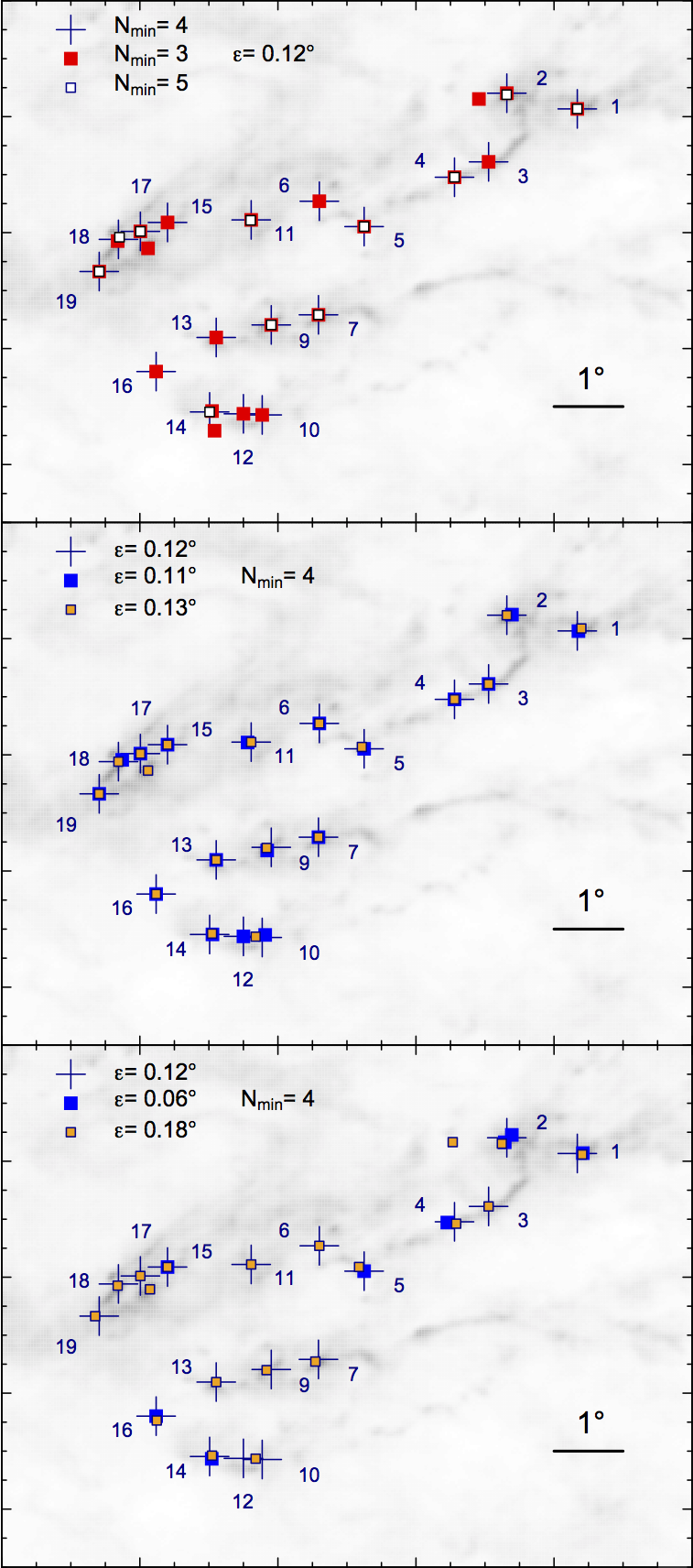}
\caption{Robustness tests of the detection of NESTs in  the three main filaments of Taurus. The  fiducial NESTs are indicated by black plus symbols and labeled in each panel. Each panel displays alternate sets of NESTs resulting from varying the selecting criteria. The  top 
panel shows the NESTs identified if $N_{min}$ is reduced to three (large red filled squares) or increased to five (small white filled squares). In the 
middle and  bottom panels, $\epsilon$ has been increased and decreased by 10\,\% and 50\,\% percent, respectively. In both panels, small orange squares and large dark blue squares represents the results of increasing and decreasing $\epsilon$, respectively.  \label{Fig:TauG_NESTs_robust}}
\end{figure}

We now focus on the NESTs robustness. Since their detection is predicated on setting the two local parameters $\epsilon$ and $N_{min}$, we must explore whether changing the values of these parameters affect the results of our analysis. The results of these tests are illustrated in Figure\,\ref{Fig:TauG_NESTs_robust}. In a first test, we have varied $N_{min}$ while keeping $\epsilon$ at its  fiducial value. Increasing this parameter by one ($N_{min}=5$) automatically eliminates all NESTs that contain only four stellar systems. However, it does not affect the detection of all others (see  top panel of Figure\,\ref{Fig:TauG_NESTs_robust}). Conversely, all  fiducial NESTs were retrieved if $N_{min}$ is decreased to 3, with the addition of three  new features, one located just  south of NEST number 17 and  the two others in the immediate vicinity of two big NESTs (numbers 2 and 14). Because they contain only three systems, these newer NESTs have a slightly  higher probability of occurring from random fluctuations, and we place a 97.9\% confidence level on their physical nature. 

We have then evaluated the effect of the local radius $\epsilon$ by varying its value by $\pm 10\%$ and $\pm 50\%$ (see  
central and  bottom panels of Figure\,\ref{Fig:TauG_NESTs_robust}) while keeping $N_{min}=4$. This range of variations is associated  with the full recovery of the multiscale stellar structure in Taurus (Joncour et al., in prep.). A decrease of ten pourcent of $\epsilon$ have had no effect on the NEST identification, while an equivalent increase lead to the identification of a new feature (containing four stars) along filament number 1, south of NESTs numbers 17 and 18, at the same location where a new feature appeared when decreasing $N_{min}$. Furthermore, the same ten pourcent increase on $\epsilon$ has driven the merging of the two NESTs numbers 10 and 12 into a single larger NEST. This is a consequence of both NESTs having the lowest values of $\alpha_N$ (see Table\,\ref{Tab:NESTsCaract}) indicating that they are only marginally separated. Increasing $\epsilon$ by 50\% did not further alter the set of identified NESTs as no other pair of NESTs are similarly poorly separated. However, decreasing $\epsilon$ by 50\%, down to $\sim 30 {\rm kAU}\,(\sim 0.15 {\rm pc})$, has resulted in only eight NESTs being identified which corresponds to the most compact "cores" of the  fiducial NESTs. The confidence level for these eight detections above random is extremely high due to their high surface density, reaching 99.997\%.

In conclusion, while the exact number and detailed properties of NESTs are dependent on the parameters $N_{min}$ and $\epsilon$, the detection of stellar NESTs in Taurus, as well as their gross properties, are robust results. In the following analysis, we have retained the 20  fiducial NESTs identified in Section\,\ref{sub:stelNESTs}. These are 
embedded in a more dispersed or hierarchically structured population on larger scales.

\subsection{NESTs in the context of previously identified groups in Taurus}

The identification of loose stellar groups in Taurus is a topic that has a rich history \citep[e.g.,][]{GomezEtAl1993,KirkMyers2011}. It is natural to compare the newly identified NESTs to these groups. In Table\,\ref{Tab:NESTsCaract}, we indicate to which group identified by \cite{GomezEtAl1993,KirkMyers2011} each NEST belongs. Five NESTs are located at the center of loose groups identified by \citeauthor{KirkMyers2011} in a one-to-one correspondence. 
On the other hand, the remaining 3 loose groups of \citeauthor{KirkMyers2011} are in fact substructured, with two or three NESTs in each. Furthermore, our analysis has revealed seven new substructures, each containing between 4 and 6 stars. 
These substructures could not have been identified by \citep{KirkMyers2011}, as these authors have used an arbitrary threshold of 10 stars per group. Nonetheless, based on our analysis, we believe that these small NESTs are physical coherent structures, given our $99.85\%$ significance level above random spatial fluctuations.

\section {NESTs properties}\label{sec:props}

\subsection{NESTs stellar content}

The NESTs contain nearly half of the Taurus stellar population, yet their total projected surface area is a small fraction of the star-forming region. Focusing first on the three central main filaments within the cloud, which appear as the main sites of ongoing star formation in Taurus, the projected area of the convex hull formed by all stars they contain is $33 \,{\rm deg}^2$ while the total projected area of all NESTs located within the three main filaments is only $0.27 \,{\rm deg} ^2$. Thus, the NESTs cover less than 1\% of the projected area of the central filaments. Expanding this analysis to the whole cloud, including the northern and southern stellar components illustrated in Figure\,\ref{Fig:TauG_NESTs_20_MST}, shows that the NESTs cover less than 0.1\% of the cloud's projected area ($0.32\, {\rm deg}^2$ out of $202\, {\rm deg}^2$).

Consequently, half of the stellar population in Taurus is concentrated in tiny, high density pockets of stars, with stellar densities that range from 50 to 2500\,stars\,pc$^{-2}$, with a median density that is $\approx100$ times higher than the average density in the whole cloud. This range of stellar surface densities places all NESTs above the median surface densities in nearby star-forming regions \citep{BressertEtAL2010}, including Ophiuchus and Orion. Indeed, the densest NESTs are close to the maximum found in these star forming regions, although it must be noted that their analysis is unable to probe densities exceeding $\approx1000$\,stars\,pc$^{-2}$.

\subsection{NESTs spatial distribution and geometry}

To further study the nature of NESTS, we have computed their "ellipsoid hull" which are  the ellipsoid of minimal area such that all given stars within the NEST lie just inside or on the boundary of the ellipsoid. These ellipses are characterized by their semi-major and semi-minor axis ($a$, $b$), eccentricity ($e$), centroid ($\alpha_C$, $\beta_C$) and major axis position angle (PA). These quantities are listed in Table\,\ref{Tab:NESTsCaract} and allows 
other studies of NESTs,  such as the cumulative distribution of their position angles,  the spatial distribution of their centroids, and derivation of their most probable  intrinsic 3D geometrical properties; all 3 topics that will be described in the following.

\begin{figure*}[!ht]
\includegraphics{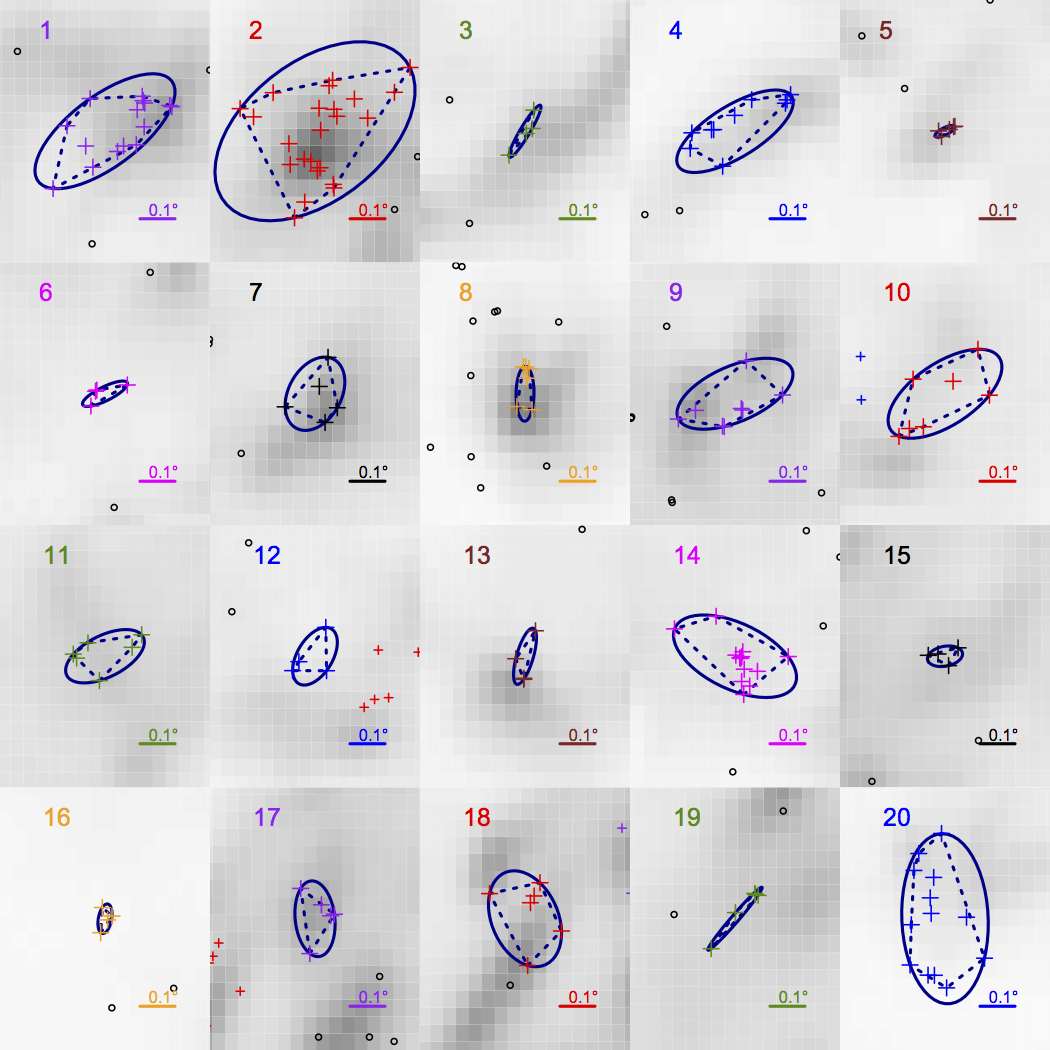}
\caption{Convex hull (dashed line) and spanning ellipsoid hull fitted on the NEST. The colored horizontal line at the bottom of each NEST scales as 0.1 degree.
\label{Fig:TauG_Extinct_NESTsBaryFil}}
\end{figure*}

\begin{figure}[!ht]
\includegraphics[width=\columnwidth]{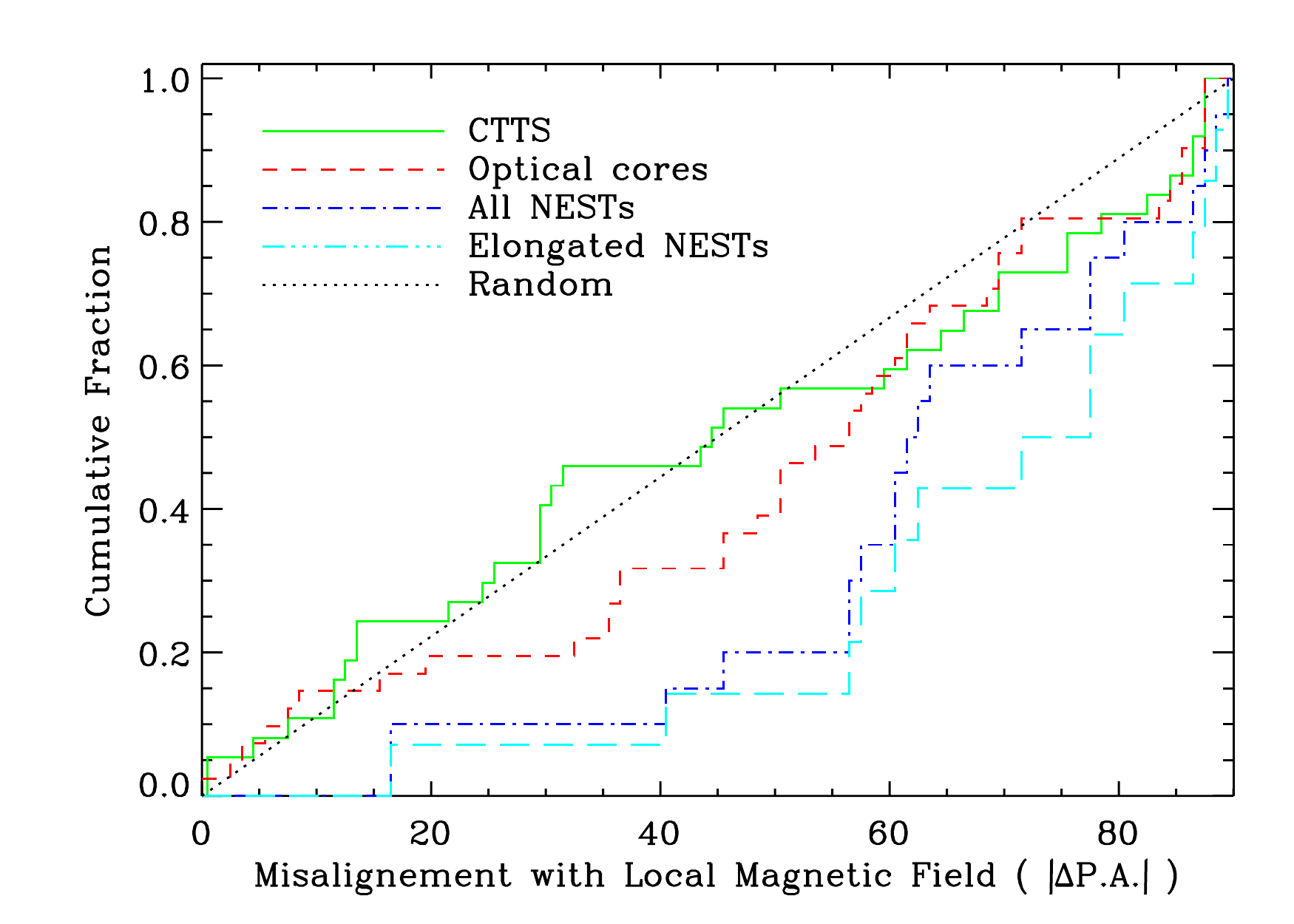}
\caption{Cumulative distribution function of the difference in position angle between the local magnetic field and the NESTs semi-major axis (dashed-dotted blue line). The long dashed light blue histogram corresponds to the subset of NESTs whose semi-major axis is more than twice as large as their semi-minor axis and, thus, whose position angle is best defined. The green solid and red  hatched histograms are the corresponding distribution for the symmetry axis of  YSO disks and dense cores, respectively \citep{MenardDuchene2004}. The black dotted line is the function expected for a randomly oriented sample.   \label{Fig:cdf_cores_nests}}
\end{figure}

As is evident in Figure\,\ref{Fig:TauG_Extinct_NESTsBaryFil}, all but two of the NESTs appear tightly concentrated along the three main filaments of the Taurus molecular cloud. Considering both the location and elongated geometry of the NESTs, we have noticed in Figure\,\ref{Fig:TauG_Extinct_NESTsBaryFil} that the NESTs appear preferentially oriented along the gas filaments.  Beyond this visual inspection, we wanted to go further and present a quantitative argument. To study the relative alignment of the NESTs, we have used the orientation of the local magnetic field traced by linear polarization measurement of background stars as reference.   Based on observations, gas filaments in molecular cloud run rather perpendicular to that local direction \citep{ChapmanEtAl2011,PlanckEtAdeEtAl2016} even if in denser environments, the orientation may be either parallel or perpendicular \citep{LiEtAl2013,ZhangEtAl2014}, although a much more complicated picture has been recently obtained in massive star forming regions \citep{KochEtAl2018}. These findings may be understood in the framework of a recent theoretical work
showing that these two configurations (i.e., at low gas column density, magnetic field tends to
be orthogonal to the density gradients, while it
tends to be parallel to them at high gas column density), are shown to be the two preferred modes that
a turbulent magnetized gas found at equilibrium \citep{SolerHennebelle2017}. 
Taurus is well known to be  a low density environment, and as such most of the magnetic fields are perpendicular to the main filaments, probably  fed by the gas along the striations \citep{PalmeirimEtAl2013}. Figure\,\ref{Fig:cdf_cores_nests} reveals quantitatively that the position angle of the NESTs is indeed preferentially oriented perpendicular to the local magnetic field, along the filaments. In comparison, \cite{MenardDuchene2004} have shown that the symmetry axis of individual  YSO disks  in Taurus is randomly oriented relative to the magnetic field, while the major axis of dense cores is intermediate between these two populations, neither completely random nor almost always perpendicular to the magnetic field \citep{LeeMyers1999,MenardDuchene2004}.  This study shows quantitatively the very close connection of the NESTs and the gas filament.

We have then studied the spacings between NESTs by themselves using the MST built from the set of the NESTs centroids (see Figure\,\ref{Fig:TauG_NESTs_20_MST}). 
We have found that the median length of a segment of the MST is 2.3\,pc. However, given the distribution of NESTs along the filaments, the distribution of segment lengths is skewed by those that connect NESTs located in different filaments. Nonetheless, the median length of the segments in the northern two filaments is 1.9--2.1\,pc (with a dispersion of $\approx$0.8\,pc), i.e., not significantly smaller. On the other hand, the three NESTs identified in filament 3  are separated by 0.7 and 1.1\,pc, respectively, significantly closer to one another. If the NESTs 10 and 12, which are located in that southernmost filament, instead constitute a single, larger NEST (see subsection \ref{Sub_sec:Robust_NESTs}),  the NESTs are distributed rather evenly along filaments, with a typical spacing of $\approx$2\,pc.

\begin{figure}
\includegraphics[width=\columnwidth]{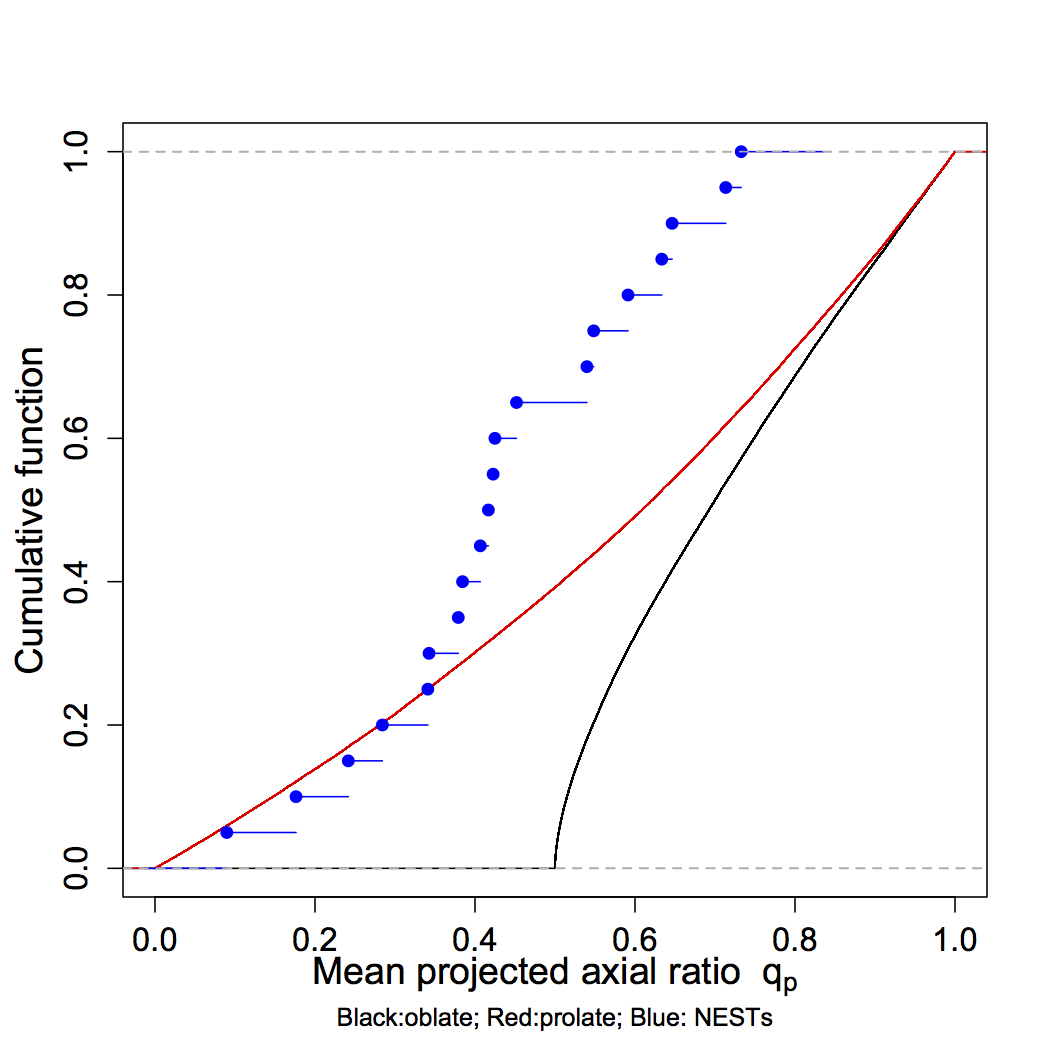}
   \caption{Cumulative distribution of the projected axial ratio $q_p$ (eq. \ref{Eq:PrOblate}) for a population of randomly oriented oblate (black) and prolate (red) ellipsoids for a fixed intrinsic (3D) axial ratio $q$. The blue solid circles mark the cumulative distribution of the 2D ratio $q_p$ of the NESTs. 
      \label{Fig:TauG_NESTs_OblaPro}}
\end{figure}

We now move on analyzing the intrinsic geometry of the NESTs. The apparent elongation of the NESTs informs their three dimensional structure, but projection effects must be taken into account to infer the latter. 
The simplest three dimensional ellipsoids shapes, such that two of three axis lengths are equal, are either prolate
(one major axis $l_a$, two same minor ones $l_b$) or oblate (two same major axis $l_a$, one minor one $l_b$). The aspect ratio $q$ is then defined as $q=l_b/l_a$. If one considers an ensemble of identical spheroids that are randomly oriented in 3D space, the relationship between the intrinsic 3D aspect ratio $q$ of these spheroids and the expected value of their projected 2D aspect ratio $q_p=b/a$ may be estimated \citep{MyersEtAl1991} as:
 \begin{equation}
   \overline{q_p}=
     \begin{cases}
        \frac{1}{2}\left ( 1 + \frac{q^2}{(1-q^2)^{1/2} }  \ln \frac{1+(1-q^2)^{1/2}}{q}\right ) & \text{(oblate case)} \\
        \frac{q}{(1-q^2)^{1/2} }  \arccos q & \text{(prolate case)}
     \end{cases}
\label{Eq:PrOblate}
\end{equation}
The expected  cumulative function of the projected aspect ratio $q_p$ (eq. \ref{Eq:PrOblate})  for randomly oriented 3D oblate and prolate ellipsoids are plotted in Figure\,\ref{Fig:TauG_NESTs_OblaPro} along with the empirical  cumulative distribution of the projected aspect ratio $q_p$ of all 20 NESTs. It is  extremely rare for oblate ellipsoids to project into a high aspect ratio structure, as this can only happen if they are observed exactly perpendicularly to their main plane. Therefore, the projected ratio $q_p$ associated to a distribution of oblate ellipsoids spans the $0.5-1$ range, excluding an oblateness hypothesis for the NESTs. The NESTs are then most probably prolate. From the Figure\,\ref{Fig:TauG_NESTs_OblaPro}, we see indeed that at small values of $q_p$, less than 0.35 (i.e., $q$ less than $\sim 0.25$, highly elongated structures), the observed blue points associated with the NESTs follow the prolate curve. But at higher values, it deviates from a prolate randomly oriented distribution, with half of the NESTs having a value of $q_p$ around 0.4, its maximal value being 0.7 ($q$ around 0.6). This results suggest that the 3D distribution of the prolate structures are not randomly oriented. It is what we expect for the NESTs structures being aligned with the gas filaments. 
It is thus tempting to assume that the NESTs are forming inside and along the gas filaments, keeping through time their pristine prolate structure despite the dynamical effects (gas expulsion, and dynamical star interaction).

\subsection{Mass and size distribution of NESTs}
\begin{figure}[!ht]
\includegraphics[width=\columnwidth]{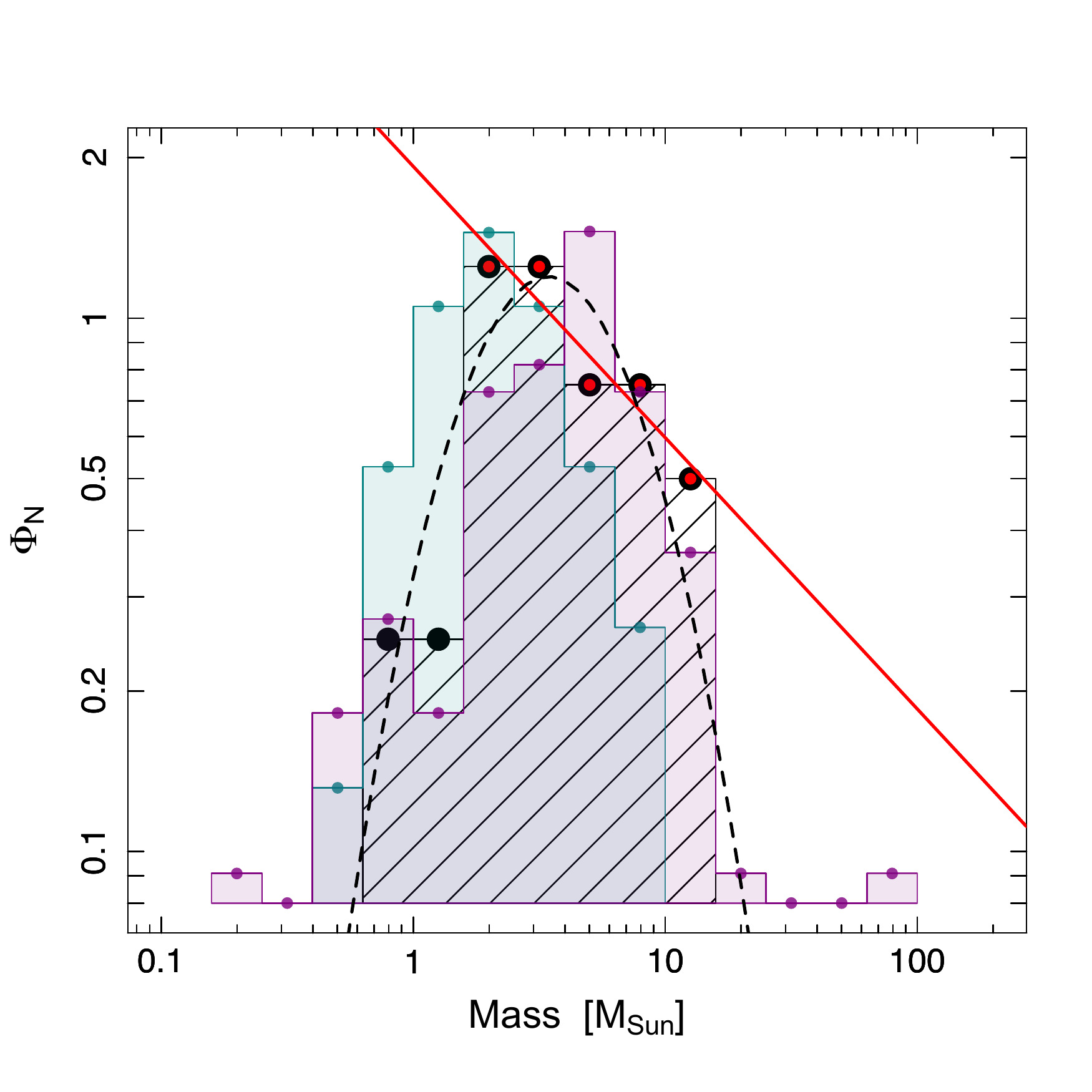}
\caption{Mass Function of NESTs (dashed histogram). The graphic shows the mass probability density function ($\Phi_N = {\rm d}N/ {d}\log m$) as a function of $\log m$, where N is the total number of the NESTs and $m$ the mass of the NESTs. For comparison, we report also  the mass functions of the H$^{13}$CO$^+$ cores \citep[][light purple]{OnishiEtAl2002} and the near-infrared extinction dust cores \citep[][light green]{SchmalzlEtAl2010}. The red curve and black dashed curve indicate respectively the power law ${\rm d}N/{d}\log m \propto m^{-\Gamma}$ fit for the high mass NESTs distribution (red points) and normal fits to the whole observed mass distribution expressed in logarithm of the mass.   \label{Fig:TauG_NEST_MassFct_Fit}}
\end{figure}

 We now study the distribution of the NESTs mass based on sum of the mass of the objects as given in full catalog of Paper\,I. It should be noted here that the reported mass of a NEST is evaluated by the sum of the primary masses of individual object, as given in Paper\,I,  as masses for stellar companions are incomplete at this stage. It is then a lower mass estimation for the mass of the NESTs. The distribution of the NESTs' mass  is shown in Figure\,\ref{Fig:TauG_NEST_MassFct_Fit} as $\Phi_N = {\rm d}N/ {d}\log m$, where $N$ is the total number of the NESTs per logarithmic mass bin and $m$ the mass of the NESTs. It is broad and heavy-tailed, ranging from $\lesssim1\,M_\odot $ to $\gtrsim10\,M_\odot$. In the logarithm mass space, a normal distribution provides an acceptable fit with a mean and standard deviation of $\overline{\log m} =0.54 \pm 0.07 $ and $\sigma (\log m) = 0.33 \pm 0.05$, respectively. Conversely, the high-mass end ($m \ge 2\,M_\odot$) of the distribution is well reproduced by a power law ($\log \Phi_N \propto m^\Gamma$) of index $\Gamma = -0.50\pm 0.1$. We note that this index is smaller than the Salpeter reference slope ($\Gamma_S = -1.35$) observed in stellar mass functions \citep{BastianEtAl2010}, but in reasonable agreement with the slope  reported for CO clumps \citep[e.g., $\Gamma = -0.65$ and -0.85;][]{KramerEtAl1996, HeithausenEtAl1998} and more recently for massive clumps \citep[$\Gamma = -0.32$, albeit at higher masses than  NESTs;][]{LiuEtAl2012}. This analysis and this connection made with the gas clump should however be taken with caution due to the low numbers statistics, as indicated by the fairly large uncertainties reported in Figure\,\ref{Fig:TauG_NEST_MassFct_Fit}.

\begin{figure*}[!ht]
\includegraphics[width=\textwidth]{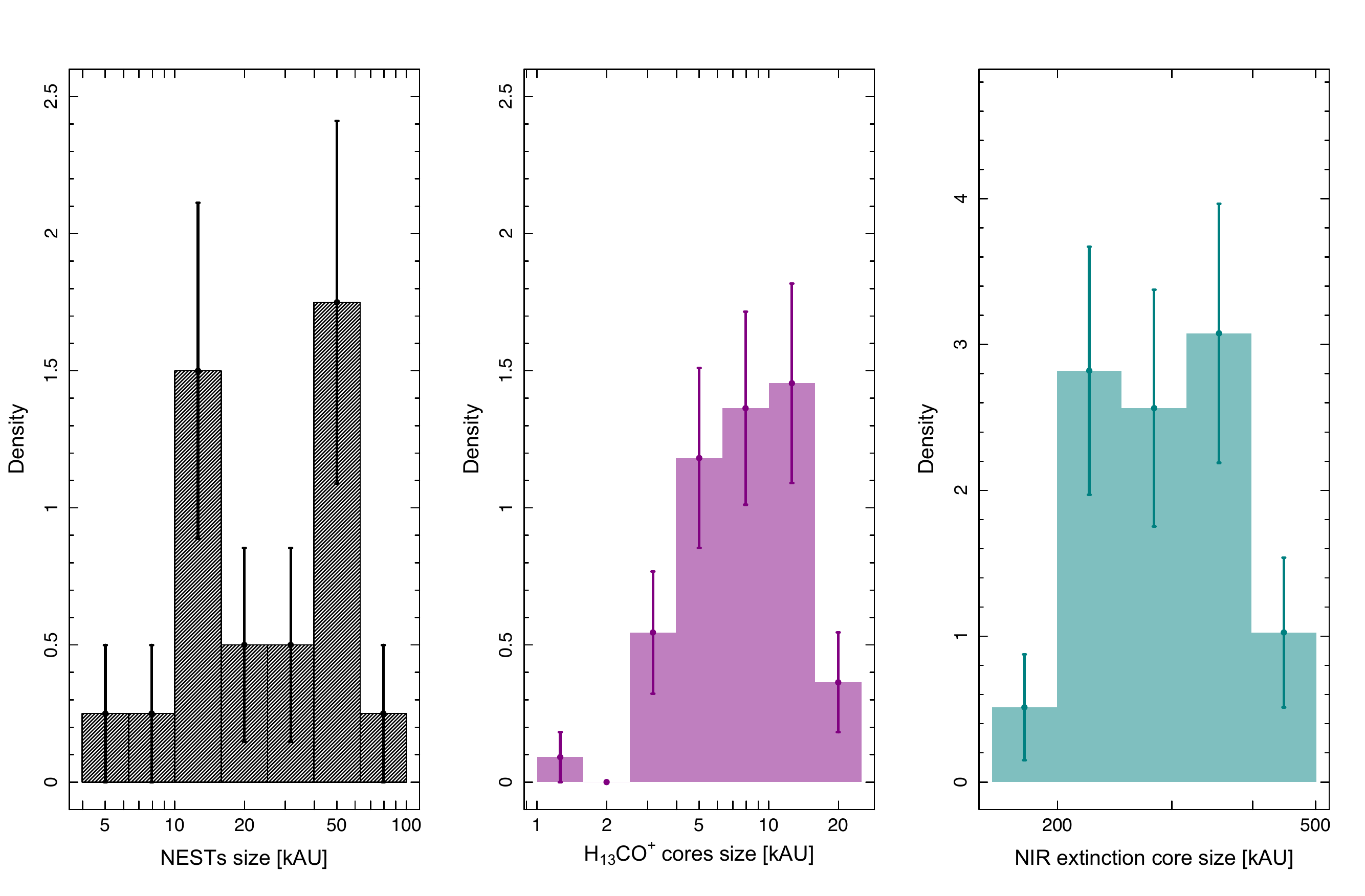}
\caption{Size distribution of NESTs (which exhibits two peaks, at 12.5 and 50 kAU, respectively; left panel), H$^{13}$CO$^+$ cores \citep[][central panel]{OnishiEtAl2002} and near-infrared extinction dust cores \citep[][right panel]{SchmalzlEtAl2010}.
   \label{Fig:TauG_NEST_Radius}}
\end{figure*}

\begin{figure}[!ht]
\includegraphics[width=\columnwidth]{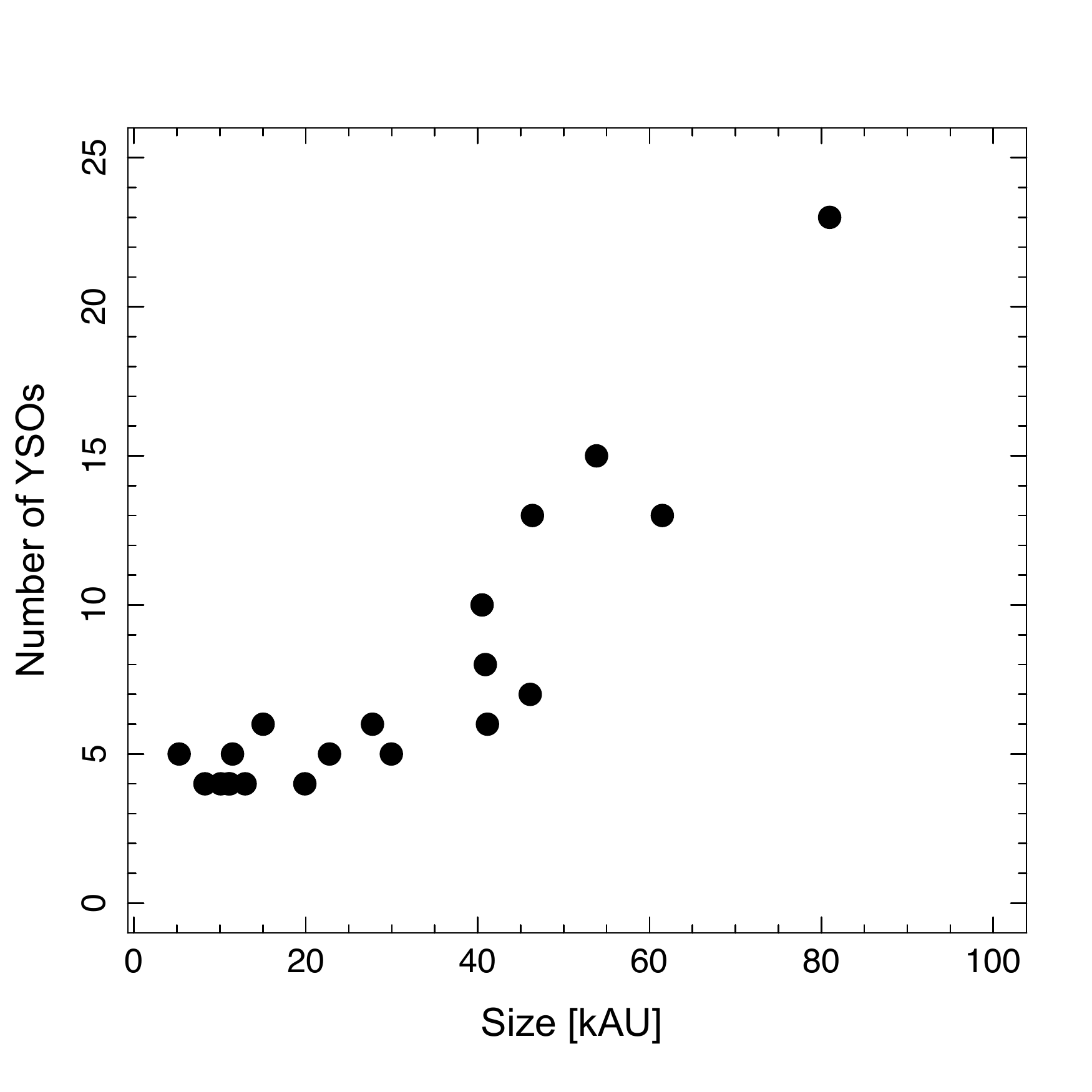}
\caption{Number of YSOs in NESTs as a function of their size.
   \label{Fig:TauG_Size_Nbre}}
\end{figure}

\begin{figure}[!ht]
\includegraphics[width=\columnwidth]{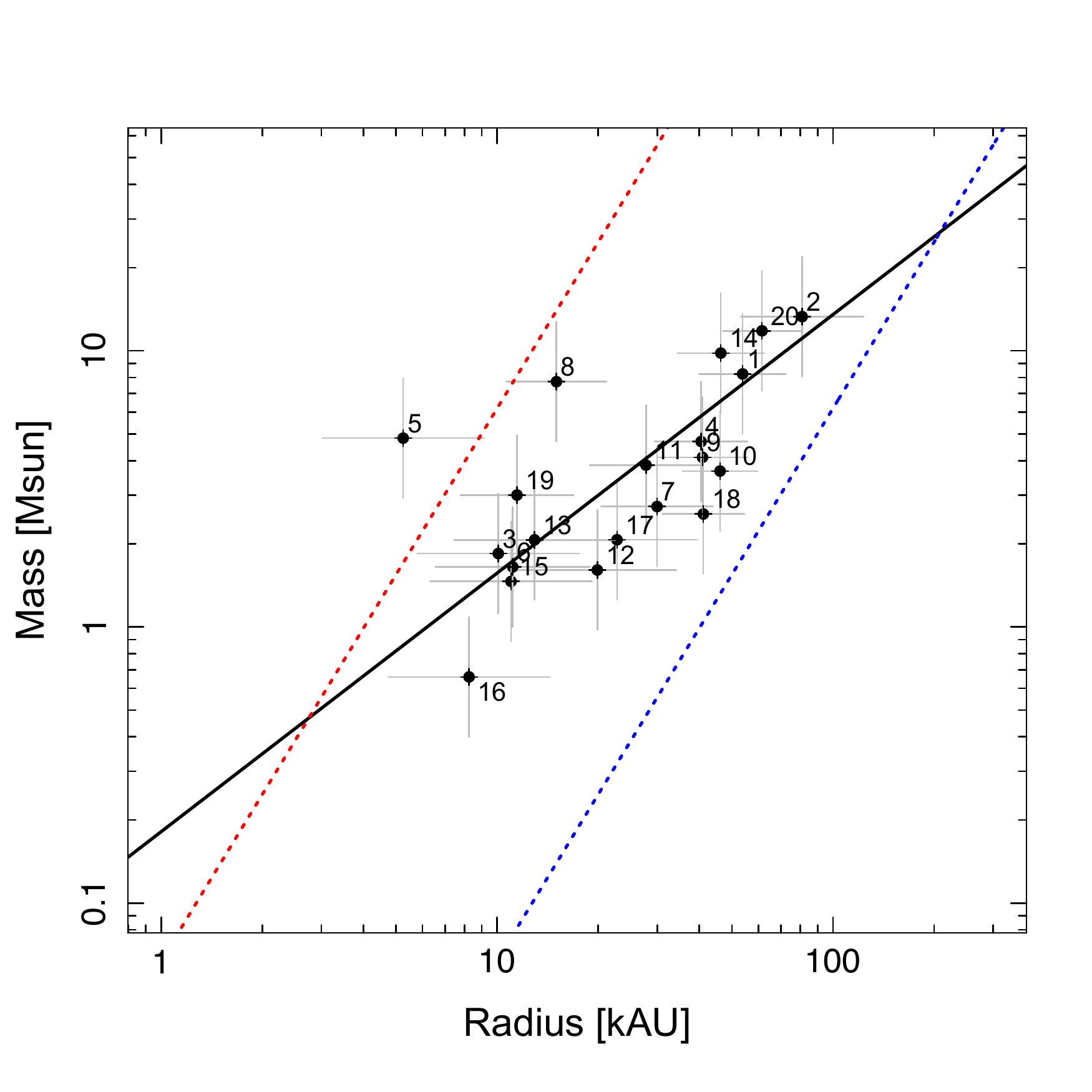}
\caption{Mass-radius relation for NESTs (black symbols labeled by NEST number). The solid black line is the best-fitting power law fit to all data points while dashed lines represent lines of constant column density ($3\,10^{22}$ cm$^{-2}$: red line; $3\,10^{20}$ cm$^{-2}$: blue line).    \label{Fig:TauG_NEST_Mass_Radius}}
\end{figure}

Contrary to the mass distribution, the size distribution of NESTs is clearly bimodal (see left panel of Figure \,\ref{Fig:TauG_NEST_Radius}). The two peaks occur at  $\approx12.5$\,kAU and $\approx50$\,kAU. Moreover, we have found that the number of YSOs inside each NEST is dependent on the NEST size, possibly revealing two distinct regimes, as illustrated in Figure\,\ref{Fig:TauG_Size_Nbre}. All the smallest NESTs ($\lesssim 30$\,kAU) contain 4 to 6 systems, whereas larger NESTs are characterized by a steady increase in the number of members with the NEST's size.
While we have found no correlation between the NESTs' radius and their stellar density, there is a positive correlation with their total stellar mass (see Figure\,\ref{Fig:TauG_NEST_Mass_Radius}). A Pearson correlation test indicates that this correlation is highly significant ($10^{-5}$ false alarm probability). To quantify this correlation, we perform a power law ($m \propto r^\gamma$) Deming fit that takes into account uncertainties on both independent quantities. Uncertainties on the radius of the NESTs are taken to be the difference between the radius of the convex hull ($R_H$) and the geometrical radius computed from the minimum spanning ellipsoid fit ($R_G =\sqrt{ab}$). The uncertainty on the total mass is dominated by uncertainties on the individual stellar masses, which may be as high as 50\% \citep{KirkMyers2011}. We havve adopted this conservative estimate in our analysis. From the Deming fit, we derive $\gamma=0.94\pm 0.21$, i.e., a nearly linear correlation. Thus the mass-radius relationship for the NESTs is markedly shallower than that of star clusters \citep[for which the power law exponent is about 5/3,][and references therein]{PfalznerEtAl2016} and of prestellar clumps and cores in the simulations \citep[][and references therein]{LeeHennebelle2016a}. On the other hand, the nearly linear behavior is reminiscent of the correlation expected for isothermal critical Bonnor-Ebert spheres, the assumed conditions setting the onset of gravitational collapse against thermal support. In that situation, the power law exponent is predicted to be exactly unity. This is consistent with observed core properties, 
 as summarized in the work of \citet[see their Figure\,7]{MotteEtAl2017}. As initially reported in \cite{MotteEtAl2001} and confirmed by \cite{KonyvesEtAl2015},  the protostellar cores that are dominated by gravity are close to the linear relation $m \propto r$, unlike the cores dominated by turbulence which have the power law $m\propto r^2$. 

\section {Discussion}\label{sec:discus}
\subsection{NESTs: Preferred sites of star formation }

Comparing the population of YSOs located inside and outside of NESTs can  give us information on the physical nature and origin of NESTs. Most importantly, Class\,I sources are the most embedded, hence likely the youngest, YSOs in Taurus. Sites of ongoing star formation are therefore expected to host a high fraction of Class\,I sources. Table\,\ref{Tab:ContingTab} summarizes the classification of all objects inside and outside of NESTs.

\begin{table}[!h]
\centering
\begin{tabular}{c|ccc|c}
 \hline 
 & Class I & Class II &  Class III  & Total\\ 
\hline
IN NEST	&  28  &75			& 48  & 151\\ 
OUT NEST & 11  & 100			&76   &187 \\  
\hline
Total & 39 & 175			& 124 & 338\\
\hline
\end{tabular}
\caption{Classification of YSOs inside (IN) and outside (OUT) the NESTs. 
\label{Tab:ContingTab}}
\end{table}

The relative proportions of Class\,II and, III sources inside the NESTs is not significantly different from that of the more dispersed population, despite a slightly lower proportion of Class\,III sources. On the other hand, the NESTs host a proportion of Class\,I sources in Taurus that is $\approx 3$ times higher than the stellar population outside of NESTs (18.5 and 5.9\%, respectively). A standard Pearson $\chi^2$ statistical test indicates that this difference is significant at the 99.9\% confidence level, in other words Class\,I sources are preferentially found within NESTs.

\begin{figure}[!ht]
\includegraphics[width=\columnwidth]{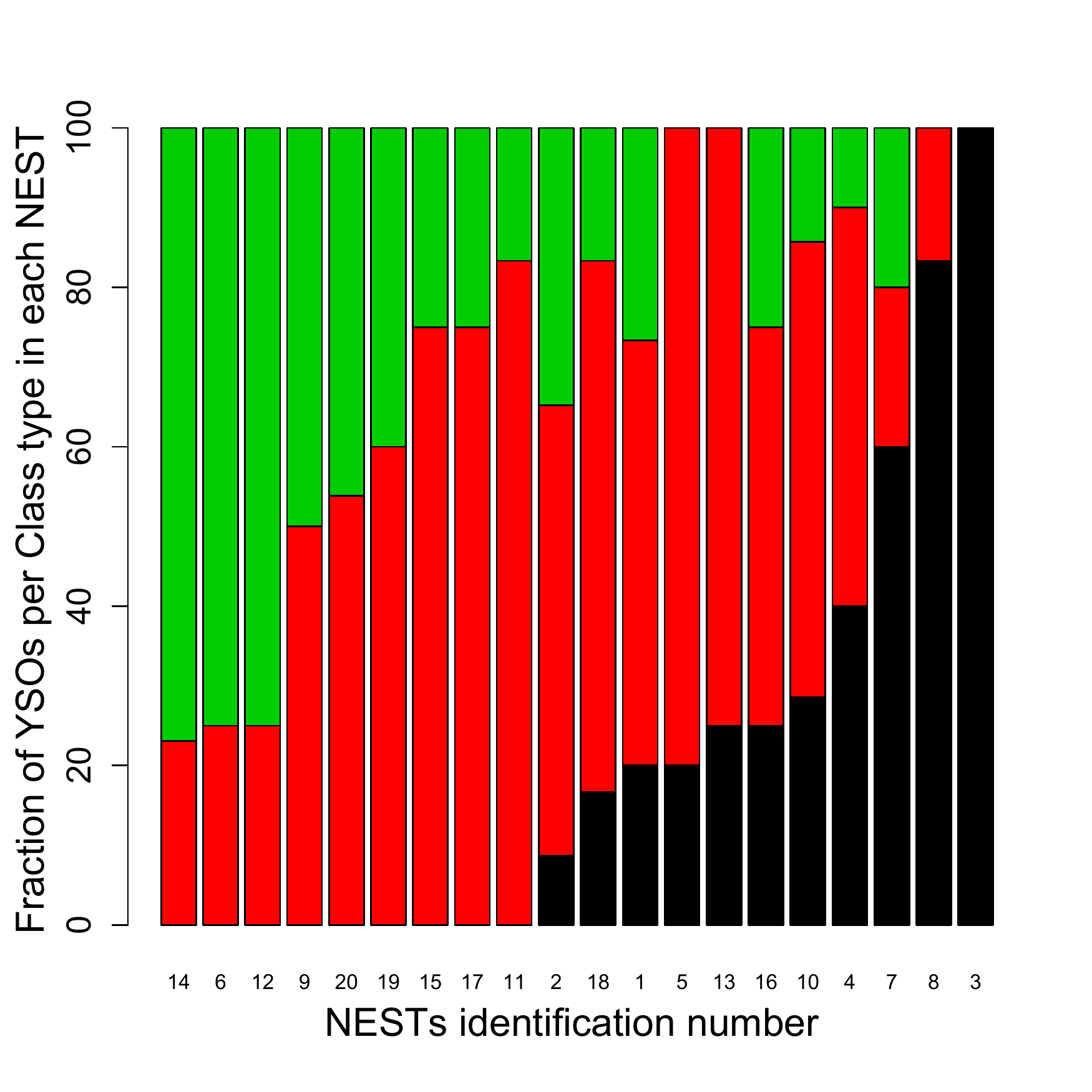}
\caption{Fraction of Class I (Black), II (Red) and, III (Green) objects within each NEST. From right to left, the NESTs are ordered by decreasing fraction of Class I (or Class II if there is no Class I  object), which could correspond to an evolutionary sequence from the younger to the older.  \label{Fig:TauG_NESTs_ClassEvol}}
\end{figure}

 Besides the fact that almost 75\% of all Taurus Class\,I sources belong to NESTs, they are concentrated in just 11 of those. This high concentration reveals that these NESTs represent regions the most presently active in Taurus. The remaining nine NESTs contain only Class II and III objects suggesting that nearly half of the NESTs are getting infertile while the other half has experienced recent star formation. Indeed, the  ratio of the number of Class I versus Class II sources and the ratio of Class II to Class III sources within NESTs suggest an evolutionary temporal scheme (see Figure \ref{Fig:TauG_NESTs_ClassEvol}). In particular, the proportion of Class I objects decreases as the proportion of Class III objects increases. When the fraction of Class III objects inside a NEST reaches 60\%, there  are no associated Class I objects, suggesting that these NESTs are the oldest. Conversely, NESTs with a high proportion of Class I objects (80\%) have no Class III objects. A NEST with such a YSOs content is thus amongst the youngest.
 
Past studies in Taurus \citep[e.g.,][]{LuhmanEtAl2010} show that Class III sources are more dispersed than Class II themselves being more dispersed than Class I objects. In our catalog, the mean 1-NNS of all Class\,I, II and III objects are 0.15, 0.34 and 0.51\,pc, respectively 
(see Table\,\ref{Tab:TauG_k1_NND_Med_Class_In_Out_All}). 
This has been interpreted in the past as a consequence of the least evolved objects being preferentially concentrated near the gas filaments and the more evolved ones being much more widely dispersed.
\begin{table}[!ht]
\centering
\begin{tabular}{r|lll}
  \hline 
  &  & 1-NNS [pc] &  \\ 
  & Class I & Class II & Class III \\
  \hline
IN NEST & 0.08 $\pm$ 0.01 & 0.09 $\pm$ 0.01 & 0.11 $\pm$ 0.01 \\ 
 ALL & 0.15 $\pm$ 0.03 & 0.34 $\pm$ 0.04 & 0.51 $\pm$0.09 \\ 
 OUT NEST & 0.30$\pm$ 0.08 & 0.52 $\pm$ 0.07 & 0.76 $\pm$ 0.14 \\ 
   \hline
\end{tabular}
\caption{Mean 1-NNS of class I, II and III for YSOs located within NESTs, outside the NESTs and for all of them. 
\label{Tab:TauG_k1_NND_Med_Class_In_Out_All}}
\end{table}
The identification of NESTs as physical substructures within Taurus prompts us to revisit this finding. For all classes, objects located outside the NESTs have larger mean 1-NNS than the overall population and these mean 1-NNSs show a marked increase  from Class\,I to Class\,III. This is in line with the previous interpretation of a dispersed population of Class\,III sources throughout the cloud. However, the mean 1-NNS for objects located within NESTs shows a qualitatively different behavior. Strikingly,  while there is also a slight marginal increase of the 1-NNS from Class\,I to Class\,III objects, all these values are  consistent with a single value for all the 3 Classes given their associated uncertainties (see Table \ref{Tab:TauG_k1_NND_Med_Class_In_Out_All}). This shows that, inside the NESTs, neither Class\,II nor Class\,III sources are more widely spread out than Class\,I sources, suggesting that the spacing between objects of a given class within NESTs does not depend on evolutionary stage. 

In an effort to search for signs of the expansion of NESTs, we searched for any correlation between the population and proportion of Class I, II and, III and their radius or surface density but found none. This suggests that there is no detectable dynamical expansion of NESTs at this scale, i.e., NESTs remain tightly packed as they age. In that respect, the observed NESTs may be pristine spatial imprints of the stellar distribution at birth. 
Most interestingly, the spatial distribution of the lowest mass YSOs appears different at a significant level than that of the most massive ones. Specifically, we have found that the majority of the former are found outside the NESTs, in the more distributed population. This aspect will be developed in more details in a forthcoming paper.

\subsection{NESTs as intermediate spatial scale structures}

With a typical size of 0.1--0.25\,pc, the NESTs are smaller structures than the the loose groups identified in the past in Taurus \citep[e.g.,][]{KirkMyers2011} and are more comparable to the size of the largest UWPs, which contain more than half of all sources in Taurus. By construction, traditional multiple systems are much smaller than NESTs. In Paper I, we have shown that the majority of UWPs have high order multiplicity, especially the closest ones with separation in the 1--10\,kAU range. Characterizing the connection between structures on these different spatial scales is important to assess whether they arise from a unique process, whereby NESTs and UWPs could be pristine imprints of a fragmentation cascade scenario.
\begin{figure}[!ht]
\includegraphics[width=\columnwidth]{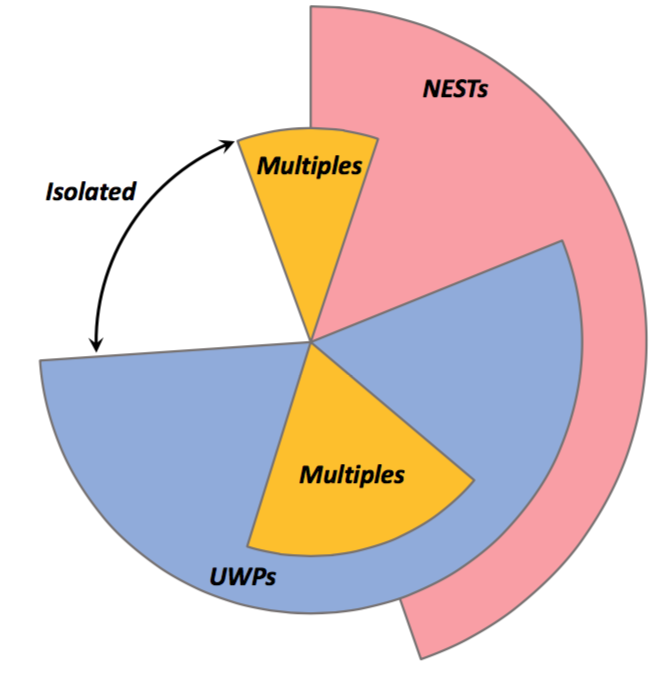}
\caption{Breakdown of the Taurus population in terms of belonging to multiple systems (separation $\leq$1\,kAU), UWPs and NESTs. The pie charts are designed to account for the fact that objects can belong to more than one category. Only $\sim$ 20\% of Taurus members are isolated single stars.
   \label{Fig:piechart}}
\end{figure}
To probe how these different scales of stellar groups are related, from the  multiple scale (5-1,000 AU) to the UWPs scale (1-60 kAU) up to the NESTs scale (15-155 kAU), we estimated the fraction of YSOs that belong in each stellar group (see Figure \ref{Fig:piechart}). The vast majority of YSOs are either members of multiples, UWPs or NESTs, leaving a  small fraction less than 20\% of single isolated  stars. Only 20\% of multiples are not included in groups of higher hierarchy  (i.e., either within UWPs or NESTs) whereas nearly half of the UWPs are inside NESTs. Only $\sim$20\% of the whole YSO population (single and multiples)  belong neither to an UWP nor to a NEST. They are isolated since they are located at the peripheries of the stellar groups and  on average 4.7 times further  away (0.56 pc, 115 kAU for the median) from their first nearest companion compared to the other YSOs that belong either to UWPs or NESTs (0.12 pc or 25 kAU for the median). Although this isolated population contains proportionally more Class III objects and less Class II and I objects, this difference does not appear statistically significant when performing the Pearson's contingency test.

We  propose to outline the connection between the different length scales through the estimate of their surface stellar density as a function of their mean radius $r$. For UWPs, the latter is estimated as half their separation, 
and the stellar surface density $\rho_*$ is then estimated as the total number of stars $n_*$ within the UWPs divided by the area, i.e., $\rho_*= n_*/(\pi r^2)$. For NESTs, loose groups \citep[as defined in][]{KirkMyers2011}, the three main filaments region and the whole Taurus cloud, the projected area is estimated from the surface $A$ of their associated convex hull containing each $n_*$ stars. We fit an ellipsoid to the convex hull and derive the semi major and minor axes (respectively $a$ and $b$). We have defined the mean geometrical radius $r$ as $r_G=\sqrt{ab}$ and the associated stellar surface density is $\rho_*=n_*/A$. 

\begin{figure}[!ht]
\includegraphics[width=\columnwidth]{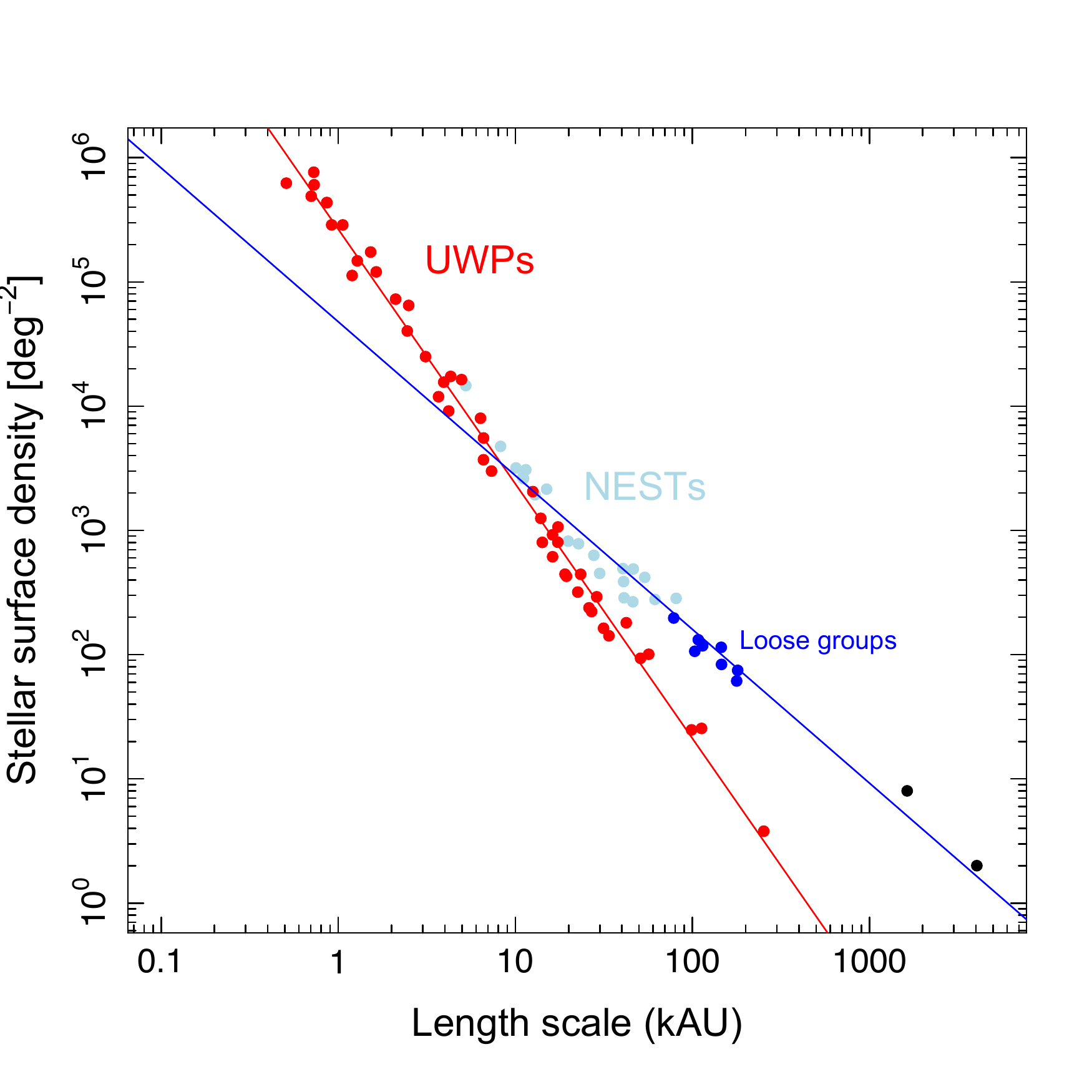}
\caption{Surface stellar density on the scale of UWPs (red dots), NESTs (light blue dots), loose groups (dark blue dots) as defined in \cite{KirkMyers2011}, the 3 main filaments as a whole and the whole Taurus  cloud (black dots). This plot outlines two main free scale regimes, respectively  associated to UWPs (red line)   and the various structures on larger scales (blue line). 
\label{Fig:TauG_Nest_WP_rho_size}}
\end{figure}

The surface stellar density of these spatial features, shown in Figure\,\ref{Fig:TauG_Nest_WP_rho_size}, reveals two regimes: one is associated with  UWPs ($\rho_* \propto r^{-2}$), and the second  with loose groups and spatial features on larger scales ($\rho_* \propto r ^{-1.2}$). The former is related to cascade fragmentation (Paper I), while the latter is associated  with a  clustering regime. NESTs appear as intermediate structures, with smaller (resp. larger) ones following  the relationship observed for UWPs (resp.loose groups and larger structures).

\subsection{ Origin of NESTs }
 
\begin{figure*}[!ht]
\includegraphics[width=\textwidth]{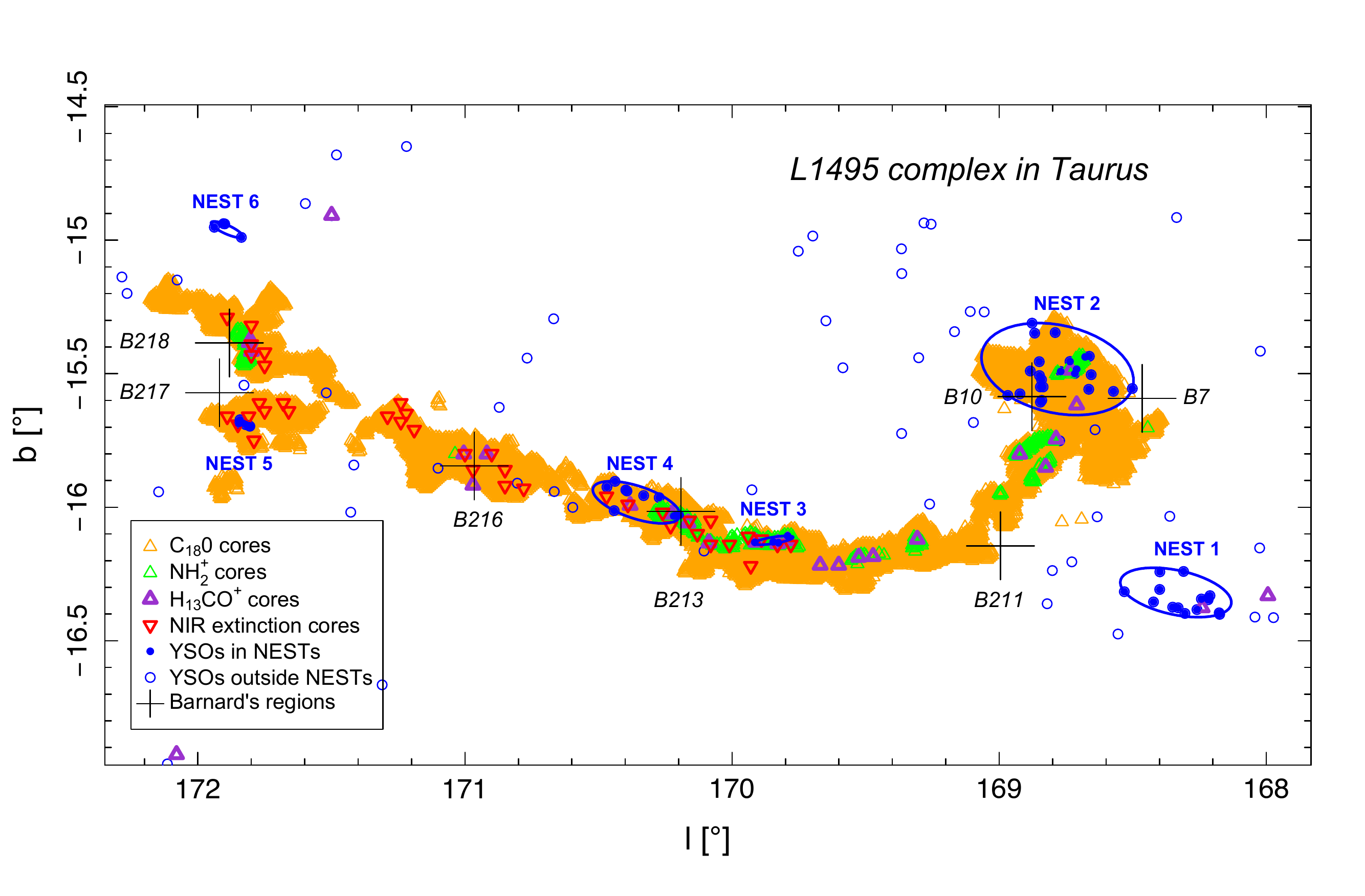}
   \caption{Stellar NESTs  1--6 and dense molecular cores in L1495 Taurus complex. Orange and green triangles indicate  C$^{18}$O and N$_2$H$^{+}$ cores \citep{HacarTafalla2013}; purple triangles mark H$^{13}$CO$^{+}$ cores \citep{OnishiEtAl2002}; finally, red triangles represent NIR dust extinction cores \citep[][we note that in their work, observations are limited to the east portion of the region, no NIR data are reported within B211, B10 and B7  regions]{SchmalzlEtAl2010}. Based on the data published in \citet[Table 1]{KenyonEtAL2008}, the center of obscured Barnard regions are plotted as black plus marks. The YSOs are plotted as solid (resp. empty) blue circle when they are inside (resp. outside) the NESTs. The NESTs' limits are shown as blue ellipsoids. We note that the L1495 complex shelter the NESTS 2--5 along the filament. The NESTs 1 and 6 are located outside the region of molecular gas studies.
 \label{Fig:TauG_gas_NESTs}}
\end{figure*}

The close connection between the NESTs and the  gas along the Taurus filaments raises the question of their origin (see Figure \ref{Fig:TauG_Extinct_NESTsBaryFil}). We have distinguished, as generally accepted, the more dense cores with typical density of $10^4-10^5$ cm$^{-3}$ and length scale of $0.05-0.1$pc from larger diffuse gas clumps with typical density of $10^3-10^4$ cm$^{-3}$.  Zooming in the L1495 Taurus region and gathering data from  NIR dust extinction and C$^{18}$O clumps \citep{SchmalzlEtAl2010,HacarTafalla2013} and dense  N$_2$H$^{+}$ and H$_{13}$CO$^{+}$ cores\citep{HacarTafalla2013,OnishiEtAl2002}, we indeed note that the large scale structure of filament as outlined by the C$^{18}$O gas tracer is closely associated with the location of both NESTs and molecular cores (see Figure \ref{Fig:TauG_gas_NESTs}). We note also  that there are no  NEST in regions where dense cold molecular cores are the most present and clustered (B216, B218, B210, B10). The exceptions are at the  edges of both N$_2$H$^{+}$  B213 clustered cores regions (NESTs number 3 and 4) and for the largest NEST (NEST number 2) located in the B7 region, where there are few dense  N$_2$H$^{+}$ cores in the central part. These three situations may illustrate different evolutionary states for the formation of stars in those regions, the latter being the most evolved.

The range of surface density in NESTs is close to that of dense cores in molecular clouds with typical H$_2$ surface density of $10^{20}$ to $10^{22}$ g/cm$^2$ (see figure \ref{Fig:TauG_NEST_Mass_Radius}). Moreover, we  have gathered  mass information on NIR dust extinction clumps \citep{SchmalzlEtAl2010} and H$^{13}$CO$^{+}$ cores \citep{OnishiEtAl2002}. While a KS test indicates that the two associated core mass distributions differ from one another (at the 99.8\% confidence level), the mass distribution of NESTs is indistinguishable with either (p-value of the same parent distribution of 0.8 and 0.1 for NIR or H$^{13}$CO$^{+}$ cores, respectively). 

These clues tend to suggest that NESTs are the direct descendants of cores, in the sense that each NEST could be connected on a one-to-one basis to a dense core that subsequently collapses and fragments. Indeed, observations showed  multiple and wide pair Class 0 or I objects  within a single core \citep{PinedaEtAl2015,SadavoyStahler2017}, indicating that several objects may form within a single core. 
However, while this relation is plausible for the small NESTs having 4-6 stars, this scenario appears less credible to explain the formation of the richer NESTs sheltering up to more than 20 stars from one single core. Indeed, to form these richer NESTs would require denser cores. Since such cores are not observed in Taurus \citep{HacarTafalla2013,OnishiEtAl2002}, a scenario implying aggregation of cores  is necessary to provide an explanation for the richer NESTs.   

The observed bimodality in the distribution of NEST sizes and the two regimes associated with their size-star number relation suggest there  are two mechanisms at  work in the formation of NESTs, allowing us to produce a coherent framework for all NESTs. The first peak in the distribution of size appears in the range of the H$^{13}$CO$^{+}$ dense core size distribution \citep{OnishiEtAl2002}, whereas the second peak is far beyond that range, nearly 4 times bigger, but this peak is also 4 times smaller than the average size of the less dense gas cores as traced by near-infrared extinction maps \citep{SchmalzlEtAl2010}. Besides, the mass distribution of the NESTs spans the same range of both mass cores. 

It is therefore tempting to associate the first regime of NESTs (those with few members)  with the fragmentation of a single core and the second regime (the richer ones)  with  a  clustering of a few cores in a chain,  as it is outlined in the work of \cite{TafallaHacar2015}. Indeed these authors report that the distances to the nearest neighbor among the N$_2$H$^{+}$ cores in L1495 mostly lie below 0.2\,pc, 41\,kAU, exactly the range of the second peak. They also find that these  dense cores tend to cluster in linear groups of three cores on average, which they call chains. The elongated geometry of the NESTs would also be compatible with this scenario, either due to the prolate nature of the cores \citep{MyersEtAl1991} or the linear fragmentation of the filament given the chain of cores as observed by \cite{TafallaHacar2015}. The process of hierarchical fragmentation from the cloud to protostars has been recently outlined in the Perseus complex based on the different gas tracers, in particular two different paths for cloud fragmentation  \cite[isolated cores and cores in filaments,][] {PokhrelEtAl2018}. We propose that the present study outlines the stellar structures resulting from this hierarchical fragmentation. Viewing the NESTs as the intermediate spatial structures between the UWPs and the loose groups, part of the UWPs should also result from this clustering scenario of cores  as proposed by \cite{Tokovinin2017}.
Furthermore, the power law fit at the high mass range of the NESTs $\Gamma=-0.5$ is  much shallower than the power law found for dense cores in Taurus, as \cite{SadavoyEtal2010} \citep[respectively][]{SchmalzlEtAl2010} found $\Gamma = -1.22$ (resp. $\Gamma = -1.2$) for the starless dense cores (resp. NIR extinction clumps) more massive than 2$ M_\odot$. But close to the the power law associated with larger CO clumps ($\Gamma = -0.65 $ (resp. $\Gamma = -0.85 $), \cite{KramerEtAl1996, HeithausenEtAl1998}) that may breakdown in smaller pieces to form spatial substructures such as cores in  a later stage of their evolution.

\subsection{Links to star formation models}

Young stars form on a wide range of scales out of their molecular parental cloud, producing aggregates, groups, clusters and distributed population  with various degrees of clumpiness and stellar density. But whether these star forming regions form as a result of a slow collapse or contraction scenarios of clumps at quasi-equilibrium  over several free-fall dynamical times 
\citep[e.g.,][]{KrumholzEtAl2006,TanEtAl2006,FariasEtAl2017}
 or they collapse more quickly (intermediate or short timescales) on the order of 1-2 dynamical time
\citep[e.g.,][]{Ballesteros-ParedesEtAl1999,ElmegreenEtAl2000,HartmannEtAl2001,BateEtAl2003}  
 mediated either by large-scale accretion flows along filaments \citep{Myers2009,SmilgysBonnell2017} and/or a global collapse \citep{HartmannBurkert2007,VazquezEtAl2017} 
 remains an open question. 

The possible diagnostics to discriminate between these models include the morphologies of subclusters, the degree of clustering and hierarchy, and the spatial distribution of stars. We intend to develop in our next paper in this series the analysis of the subclusters hierarchy in Taurus to expand in that direction. 
For now, focusing on the densest part of Taurus, i.e., NESTs, may already prove informative. In the slow type models, the morphology of subclusters tend to be rather spherically symmetric \citep{TanEtAl2006} in contrast with the rapid type model in which elongated structures dominate \citep{BateEtAl2003}. Thus, the highly elongated NESTs we found tend to favor the rapid scenario. This is in agreement with \cite{HartmannEtAl2001}, who have made the assumption of a rapid formation and dissipation of molecular clouds in a few dynamical times to explain the small age spread (1-3 Myr) of the Taurus members and the absence of the Post T Tauri stars. However in all scenarios, the most recent works  show that the subclusters that form at local scale merge afterwards within larger ones that themselves merge to finally join a large central cluster fueled with gas and stars. But if the mixed stellar populations we observed within NESTs were to be due to the merging of previous physically unrelated sub-units, we do not expect to see an evolution of the status of the population inside the NESTS  as we have noted. An important caveat is that most of these models, and specially the slow type models, have been developed to simulate massive and dense star forming regions, and cannot be directly compared to the Taurus region. To our knowledge, no specific model to explain the formation and the persistence of the NESTs in Taurus has been proposed.

To conclude this section, we suggest that the NESTs are pristine imprints of stellar formation and that they are representative of two fragmentation scenarios. The first one is associated  with the fragmentation of a single dense core that could give birth to a small number of stellar systems (4-6), while the second one corresponds to the fragmentation of a filament in a few  (2-5) cores closely spatially associated, and thus spatially clustered. These structures may remain mostly unchanged for at least a few millions years.

\section{Conclusion}\label{sec:conclu}
Building on the stellar spatial distribution in the Taurus star-forming region, we have used the density based clustering algorithm {\tt dbscan} to identify local overdense subclusters. We have set the two free parameters of the algorithm based on the one-point correlation function and the cumulative function  of the k-nearest neighbor separations, in order to reach a 99.8\% significance level of detection above random. We found 20 stellar structures that we dub NESTs (Nested Elementary STructures). We have then performed a set of statistical studies  to derive their main properties.

Our work has shows  that about 45\% of the whole stellar population in Taurus  is concentrated in 20  small and most probably prolate  (median eccentricity of 0.9) NESTs that are regularly spaced ($\approx 2$ pc)  and mainly oriented along the principal gas filaments axes. Five of these NESTs are associated  with the densest parts of  eight previously  identified  loose groups \citep{GomezEtAl1993,KirkMyers2011}. Each of the 3 remaining loose groups are  composed  by two or three stellar NESTs and seven  of the NESTs   were not identified so far. 
The stellar NESTs contain between 4 to 23 stars each, the median being  5-6. Inside the NESTs, the surface density of stars may be as high as 2500 pc$^{-2}$ \citep[while the mean surface density of the Orion Nebula Cluster is about 1,000 YSOS/pc$^2$,][]{BressertEtAL2010}. The mean value is of the order of 340 pc$^{-2}$, which agrees well with the initial substructures density (350 pc$^{-2}$) required to reproduce the present-day binary properties in Taurus when starting from a universal initial binary distribution according to Nbody numerical simulations  \citep{MarksKroupa2012}.
 
 Although the proportion of Class\,II and, Class\,III objects inside the NESTs is not significantly different from that of the more dispersed population, the NESTs as a whole contain 3/4 of the class I objects. Studying them individually,
only half (11)  of  these NESTs contain those class 0 or I objects,  showing that they are the sites of privileged  star formation. The balance between  Class\,I, II and, III  fraction  within the NESTs suggests that they may be ordered as an evolutionary temporal scheme, some of them getting infertile with time, while other still giving birth to young stars. The fact that Class\,III objects are still present in such  a relatively tight environment suggests that they may get old and remain in the same environment for few millions years. Furthermore, the higher mass stars of the population are equally found inside and outside the NESTs, but the great majority (60-80\%) of their lowest mass  counterparts are found outside the NESTs.  This specific point will be developed in a forthcoming paper. The  total mass of stellar NESTs ranges from half a solar mass up to 10\,$M_{\odot}$, with a median mass of 3\,$M_{\odot}$. The size distribution of the NESTs is bimodal with one peak at   12.5\,kAU and the second at 50\,kAU. 

We have identified the preferred sites of star formation in Taurus as the densest stellar groups of the region. Each NEST may be the individual stellar outcome of the gravitational
collapse of a cloud that fragments to give groups of stars within few millions years.
Cloud fragmentation being prior to the formation of the YSOs, it could lead to the observable  clustering of dense cores and then stars. These NESTs are intermediate structures between the  ultra-wide pairs and the loose groups, and provide an explanation for the elbow observed in the two-point correlation function and mark the transition between the multiplicity regime and larger scale  structures.  The study of the relationship between the different scales of stellar groups, NESTs, UWPs and multiple systems, reveals that they are inter-connected, with only 20\% of stars truly single and isolated.

We propose that the youngest NESTs  are the spatial imprints of the stellar distribution as they may have emerged from their natal cloud at a scale that have been overlooked up to now in Taurus.  Using DR2 GAIA release should provide invaluable information both on distance to probe the 3D structure and on kinematics  to further probe the dynamical status of these features and their origin.  We also intend in future work to analyze other star forming regions (IC348, Orion, Chamaleon, $\rho$ Ophiucus, ...) to compare the stellar substructures to highlight similarities and differences.

\begin{acknowledgements}
We gratefully thank the anonymous referee for the careful reading and helpful suggestions to improve the paper. We also thank Lee Mundy for helpful discussions
and Marc Pound for reading the draft of this article. This work was funded  by the European Commission through the  EU H2020 Framework Program for Research and Innovation H2020-COMPET-2015-RIA-687528.
\end{acknowledgements}

\bibliographystyle{aa} 
\bibliography{Paper_NEST_Taurus_refs} 

\appendix
\section{Catalogs of stars in NESTs}
\label{AppSec:StarsInNests}

Table\,\ref{tab:NESTstars} lists all Taurus members that are located within NESTs. Details on the multiplicity of each systems (separations and references) can be found in Table\,C.1 of Paper\,I.

\longtab{
\onecolumn
{\scriptsize
\begin{longtable}{rrllrrlrlrlllll}
  \caption{Members of Taurus belonging to NESTs \label{tab:NESTstars}}\\
  \hline
 &  \# & 2MASS & Name &  RA$\_{J2000}$ &  DEC$\_{J2000}$&SpT   & M [$M_\odot$] & Class & n$_{*}$ \\  \hline
  \hline
  \endfirsthead
\caption{continued.}\\
\hline\hline
  &\# & 2MASS & Name &  RA$\_{J2000}$ &  DEC$\_{J2000}$&SpT   & M [$M_\odot$] & Class & n$_{*}$ \\  \hline
\endhead
\hline
1 & 1 & J04135328+2811233 & IRAS04108+2803A & 63.472 & 28.19 & M4c & 0.271 & I & 1 \\ 
  2 & 1 & J04135471+2811328 & IRAS04108+2803B & 63.478 & 28.19 & M2c & 0.575 & I & 1 \\ 
  3 & 1 & J04141188+2811535 & - & 63.550 & 28.20 & M6.25 & 0.086 & II & 1 \\ 
  4 & 1 & - & IRAS04111+2800G & 63.551 & 28.14 & M2c & 0.575 & I & 1 \\ 
  5 & 1 & J04141291+2812124 & V773TauA+B & 63.554 & 28.20 & K3 & 1.796 & II & 3 \\ 
  6 & 1 & J04141358+2812492 & FMTau & 63.557 & 28.21 & M0 & 0.701 & II & 1 \\ 
  7 & 1 & J04141700+2810578 & CWTau & 63.571 & 28.18 & K3 & 1.796 & II & 1 \\ 
  8 & 1 & J04141760+2806096 & CIDA1 & 63.573 & 28.10 & M5.5 & 0.137 & II & 1 \\ 
  9 & 1 & J04142639+2805597 & MHO2 & 63.610 & 28.10 & M2.5 & 0.486 & II & 3 \\ 
  10 & 1 & J04143054+2805147 & MHO3 & 63.627 & 28.09 & K7 & 0.801 & II & 2 \\ 
  11 & 1 & J04144739+2803055 & XEST20-066 & 63.697 & 28.05 & M5.25 & 0.157 & III & 1 \\ 
  12 & 1 & J04144928+2812305 & FOTauA+B & 63.705 & 28.21 & M3.5 & 0.335 & II & 2 \\ 
  13 & 1 & J04145234+2805598 & XEST20-071 & 63.718 & 28.10 & M3.25 & 0.366 & III & 1 \\ 
  14 & 1 & J04150515+2808462 & CIDA2 & 63.771 & 28.15 & M5.5 & 0.137 & III & 2 \\ 
  15 & 1 & J04151471+2800096 & KPNO1 & 63.811 & 28.00 & M8.5 & 0.022 & III & 1 \\ 
  16 & 2 & J04173893+2833005 & LkCa5 & 64.412 & 28.55 & M2 & 0.575 & III & 2 \\ 
  17 & 2 & J04174965+2829362 & V410X-ray1 & 64.457 & 28.49 & M4 & 0.271 & II & 1 \\ 
  18 & 2 & J04180796+2826036 & V410X-ray3 & 64.533 & 28.43 & M6 & 0.096 & III & 2 \\ 
  19 & 2 & J04181710+2828419 & V410Anon13 & 64.571 & 28.48 & M5.75 & 0.116 & II & 1 \\ 
  20 & 2 & J04182909+2826191 & V410Anon25 & 64.621 & 28.44 & M1 & 0.633 & III & 1 \\ 
  21 & 2 & J04183110+2827162 & V410TauA+B+C & 64.630 & 28.45 & K7 & 0.801 & III & 3 \\ 
  22 & 2 & J04183112+2816290 & DDTauA+B & 64.630 & 28.27 & M3.5 & 0.335 & II & 2 \\ 
  23 & 2 & J04183158+2816585 & CZTauA+B & 64.632 & 28.28 & M3 & 0.398 & II & 2 \\ 
  24 & 2 & J04183203+2831153 & IRAS04154+2823 & 64.633 & 28.52 & M2.5 & 0.486 & I & 1 \\ 
  25 & 2 & J04183444+2830302 & V410X-ray2 & 64.644 & 28.51 & M0 & 0.701 & II & 1 \\ 
  26 & 2 & J04184023+2824245 & V410X-ray4 & 64.668 & 28.41 & M4 & 0.271 & III & 1 \\ 
  27 & 2 & J04184061+2819155 & V892Tau & 64.669 & 28.32 & B9 & 3.250 & II & 3 \\ 
  28 & 2 & J04184133+2827250 & LR1 & 64.672 & 28.46 & K4.5 & 1.378 & II & 1 \\ 
  29 & 2 & J04184250+2818498 & V410X-ray7 & 64.677 & 28.31 & M0.75 & 0.650 & II & 2 \\ 
  30 & 2 & J04184703+2820073 & Hubble4 & 64.696 & 28.34 & K7 & 0.801 & III & 2 \\ 
  31 & 2 & J04185115+2814332 & KPNO2 & 64.713 & 28.24 & M7.5 & 0.044 & III & 1 \\ 
  32 & 2 & J04185147+2820264 & CoKuTau/1 & 64.715 & 28.34 & M0 & 0.701 & II & 2 \\ 
  33 & 2 & J04185813+2812234 & IRAS04158+2805 & 64.742 & 28.21 & M5.25 & 0.157 & I & 1 \\ 
  34 & 2 & J04190110+2819420 & V410X-ray6 & 64.755 & 28.33 & M5.5 & 0.137 & II & 1 \\ 
  35 & 2 & J04190197+2822332 & V410X-ray5a & 64.758 & 28.38 & M5.5 & 0.137 & III & 1 \\ 
  36 & 2 & J04191281+2829330 & FQTauA+B & 64.803 & 28.49 & M3 & 0.398 & II & 2 \\ 
  37 & 2 & J04192625+2826142 & V819Tau & 64.859 & 28.44 & K7 & 0.801 & II & 1 \\ 
  38 & 2 & J04193545+2827218 & FRTau & 64.898 & 28.46 & M5.25 & 0.157 & II & 1 \\ 
  39 & 3 & J04194148+2716070 & IRAS04166+2708 & 64.923 & 27.27 & M0c & 0.701 & I & 1 \\ 
  40 & 3 & - & IRAS04166+2706 & 64.927 & 27.23 & M3c & 0.398 & I & 1 \\ 
  41 & 3 & J04194657+2712552 & [GKH94]41 & 64.944 & 27.22 & M7.5 & 0.044 & I & 1 \\ 
  42 & 3 & J04195844+2709570 & IRAS04169+2702 & 64.994 & 27.17 & M0c & 0.701 & I & 2 \\ 
  43 & 4 & J04210795+2702204 & - & 65.283 & 27.04 & M5.25 & 0.157 & II & 1 \\ 
  44 & 4 & J04211038+2701372 & IRAS04181+2654B & 65.293 & 27.03 & K7 & 0.801 & I & 1 \\ 
  45 & 4 & J04211146+2701094 & IRAS04181+2654A & 65.298 & 27.02 & M3 & 0.398 & I & 1 \\ 
  46 & 4 & J04213459+2701388 & - & 65.394 & 27.03 & M5.5 & 0.137 & II & 1 \\ 
  47 & 4 & J04214631+2659296 & - & 65.443 & 26.99 & M5.75 & 0.116 & II & 1 \\ 
  48 & 4 & J04215450+2652315 & - & 65.477 & 26.88 & M8.5 & 0.022 & III & 1 \\ 
  49 & 4 & J04220069+2657324 & Haro6-5B & 65.503 & 26.96 & K5 & 1.121 & I & 1 \\ 
  50 & 4 & J04220217+2657304 & FSTauA+B & 65.509 & 26.96 & M0 & 0.701 & II & 2 \\ 
  51 & 4 & J04221568+2657060 & XEST11-078 & 65.565 & 26.95 & M1 & 0.633 & I & 1 \\ 
  52 & 4 & J04221675+2654570 & - & 65.570 & 26.92 & M1.5 & 0.604 & II & 1 \\ 
  53 & 5 & J04265352+2606543 & FVTauA+B & 66.723 & 26.12 & K5 & 1.121 & II & 2 \\ 
  54 & 5 & J04265440+2606510 & FVTau/cA+B & 66.727 & 26.11 & M2.5 & 0.486 & II & 2 \\ 
  55 & 5 & J04265732+2606284 & KPNO13 & 66.739 & 26.11 & M5 & 0.178 & II & 1 \\ 
  56 & 5 & J04270266+2605304 & DGTauB & 66.761 & 26.09 & K2c & 2.134 & I & 1 \\ 
  57 & 5 & J04270469+2606163 & DGTau & 66.770 & 26.10 & K6 & 0.906 & II & 1 \\ 
  58 & 6 & J04292071+2633406 & J1-507 & 67.336 & 26.56 & M4 & 0.271 & III & 2 \\ 
  59 & 6 & J04294155+2632582 & DHTauA+B & 67.423 & 26.55 & M1 & 0.633 & II & 2 \\ 
  60 & 6 & J04294247+2632493 & DITauA+B & 67.427 & 26.55 & M0 & 0.701 & III & 2 \\ 
  61 & 6 & J04294568+2630468 & KPNO5 & 67.440 & 26.51 & M7.5 & 0.044 & III & 1 \\ 
  62 & 7 & J04292373+2433002 & GVTauA+B & 67.349 & 24.55 & K5 & 1.121 & I & 2 \\ 
  63 & 7 & J04293008+2439550 & IRAS04264+2433 & 67.375 & 24.67 & M1 & 0.633 & I & 1 \\ 
  64 & 7 & J04293209+2430597 & - & 67.384 & 24.52 & M3c & 0.398 & I & 1 \\ 
  65 & 7 & J04293606+2435556 & XEST13-010 & 67.400 & 24.60 & M3 & 0.398 & III & 1 \\ 
  66 & 7 & J04295950+2433078 & - & 67.498 & 24.55 & M5 & 0.178 & II & 1 \\ 
  67 & 8 & J04313407+1808049 & L1551/IRS5 & 67.892 & 18.13 & K0c & 2.430 & I & 1 \\ 
  68 & 8 & J04313613+1813432 & LkHa358 & 67.901 & 18.23 & K8 & 0.790 & I & 1 \\ 
  69 & 8 & J04313747+1812244 & HH30 & 67.906 & 18.21 & M0 & 0.701 & I & 1 \\ 
  70 & 8 & J04313843+1813576 & HLTau & 67.910 & 18.23 & K7 & 0.801 & I & 1 \\ 
  71 & 8 & J04314007+1813571 & XZTauA+B & 67.917 & 18.23 & M2 & 0.575 & II & 2 \\ 
  72 & 8 & J04314444+1808315 & L1551NE & 67.935 & 18.14 & K0c & 2.430 & I & 1 \\ 
  73 & 9 & J04315056+2424180 & HKTauA+B & 67.961 & 24.41 & M0.5 & 0.667 & II & 2 \\ 
  74 & 9 & J04321540+2428597 & Haro6-13 & 68.064 & 24.48 & M0 & 0.701 & II & 1 \\ 
  75 & 9 & J04321786+2422149 & - & 68.074 & 24.37 & M5.75 & 0.116 & III & 1 \\ 
  76 & 9 & J04321885+2422271 & V928TauA+B & 68.079 & 24.37 & M0.5 & 0.667 & III & 2 \\ 
  77 & 9 & J04323058+2419572 & FYTau & 68.127 & 24.33 & K5 & 1.121 & II & 1 \\ 
  78 & 9 & J04323176+2420029 & FZTau & 68.132 & 24.33 & M0 & 0.701 & II & 1 \\ 
  79 & 9 & J04325026+2422115 & - & 68.209 & 24.37 & M7.5 & 0.044 & III & 1 \\ 
  80 & 9 & J04330197+2421000 & MHO8 & 68.258 & 24.35 & M6 & 0.096 & III & 2 \\ 
  81 & 10 & J04322415+2251083 & - & 68.101 & 22.85 & M4.5 & 0.225 & II & 1 \\ 
  82 & 10 & J04323205+2257266 & IRAS04295+2251 & 68.134 & 22.96 & K7 & 0.801 & I & 1 \\ 
  83 & 10 & J04324911+2253027 & JH112 & 68.205 & 22.88 & K6 & 0.906 & II & 4 \\ 
  84 & 10 & J04330945+2246487 & - & 68.289 & 22.78 & M6 & 0.096 & II & 1 \\ 
  85 & 10 & J04331650+2253204 & IRAS04302+2247 & 68.319 & 22.89 & M0c & 0.701 & I & 1 \\ 
  86 & 10 & J04331907+2246342 & IRAS04303+2240 & 68.329 & 22.78 & M0.5 & 0.667 & II & 1 \\ 
  87 & 10 & J04332621+2245293 & XEST17-036 & 68.359 & 22.76 & M4 & 0.271 & III & 1 \\ 
  88 & 11 & J04330781+2616066 & KPNO14 & 68.283 & 26.27 & M6 & 0.096 & III & 1 \\ 
  89 & 11 & J04331435+2614235 & IRAS04301+2608 & 68.310 & 26.24 & M0 & 0.701 & II & 1 \\ 
  90 & 11 & J04333678+2609492 & ISTauA+B & 68.403 & 26.16 & M0 & 0.701 & II & 2 \\ 
  91 & 11 & J04334465+2615005 & - & 68.436 & 26.25 & M4.75 & 0.201 & II & 1 \\ 
  92 & 11 & J04335245+2612548 & - & 68.469 & 26.22 & M8.5 & 0.022 & II & 1 \\ 
  93 & 11 & J04335470+2613275 & ITTauA & 68.478 & 26.22 & K2 & 2.134 & II & 2 \\ 
  94 & 12 & J04335200+2250301 & CITau & 68.467 & 22.84 & K7 & 0.801 & II & 1 \\ 
  95 & 12 & J04335252+2256269 & XEST17-059 & 68.469 & 22.94 & M5.75 & 0.116 & III & 1 \\ 
  96 & 12 & J04341099+2251445 & JH108 & 68.546 & 22.86 & M1 & 0.633 & III & 1 \\ 
  97 & 12 & J04341527+2250309 & CFHT1 & 68.564 & 22.84 & M7 & 0.057 & III & 1 \\ 
  98 & 13 & J04352737+2414589 & DNTau & 68.864 & 24.25 & M0 & 0.701 & II & 1 \\ 
  99 & 13 & - & IRAS04325+2402C & 68.897 & 24.14 & M8c & 0.031 & II & 1 \\ 
  100 & 13 & J04353539+2408194 & IRAS04325+2402A+B & 68.897 & 24.14 & M0c & 0.701 & I & 1 \\ 
  101 & 13 & J04354093+2411087 & CoKuTau3A+B & 68.921 & 24.19 & M1 & 0.633 & II & 2 \\ 
  102 & 14 & J04352089+2254242 & FFTauA+B & 68.837 & 22.91 & K7 & 0.801 & III & 2 \\ 
  103 & 14 & J04354203+2252226 & XEST08-033 & 68.925 & 22.87 & M4.75 & 0.201 & III & 1 \\ 
  104 & 14 & J04354733+2250216 & HQTau & 68.947 & 22.84 & K2 & 2.134 & II & 1 \\ 
  105 & 14 & J04355109+2252401 & KPNO15 & 68.963 & 22.88 & M2.75 & 0.442 & III & 1 \\ 
  106 & 14 & J04355143+2249119 & KPNO9 & 68.964 & 22.82 & M8.5 & 0.022 & III & 1 \\ 
  107 & 14 & J04355209+2255039 & XEST08-047 & 68.967 & 22.92 & M4.5 & 0.225 & III & 1 \\ 
  108 & 14 & J04355277+2254231 & HPTau & 68.970 & 22.91 & K3 & 1.796 & II & 2 \\ 
  109 & 14 & J04355286+2250585 & XEST08-049 & 68.970 & 22.85 & M4.25 & 0.248 & III & 1 \\ 
  110 & 14 & J04355349+2254089 & HPTau/G3 & 68.973 & 22.90 & K7 & 0.801 & III & 2 \\ 
  111 & 14 & J04355415+2254134 & HPTau/G2 & 68.976 & 22.90 & G0 & 2.659 & III & 1 \\ 
  112 & 14 & J04355684+2254360 & Haro6-28A+B & 68.987 & 22.91 & M3 & 0.398 & II & 2 \\ 
  113 & 14 & J04361038+2259560 & CFHT2 & 69.043 & 23.00 & M7.5 & 0.044 & III & 1 \\ 
  114 & 14 & J04363893+2258119 & CFHT3 & 69.162 & 22.97 & M7.75 & 0.038 & III & 1 \\ 
  115 & 15 & J04381486+2611399 & - & 69.562 & 26.19 & M7.25 & 0.051 & II & 1 \\ 
  116 & 15 & J04382134+2609137 & GMTau & 69.589 & 26.15 & M6.5 & 0.076 & II & 1 \\ 
  117 & 15 & J04382858+2610494 & DOTau & 69.619 & 26.18 & M0 & 0.701 & II & 1 \\ 
  118 & 15 & J04383528+2610386 & HVTauA+B & 69.647 & 26.18 & M1 & 0.633 & III & 3 \\ 
  119 & 16 & J04385859+2336351 & - & 69.744 & 23.61 & M4.25 & 0.248 & II & 1 \\ 
  120 & 16 & J04390163+2336029 & - & 69.757 & 23.60 & M6 & 0.096 & II & 1 \\ 
  121 & 16 & J04390525+2337450 & - & 69.772 & 23.63 & M4c & 0.271 & I & 1 \\ 
  122 & 16 & J04390637+2334179 & - & 69.777 & 23.57 & M7.5 & 0.044 & III & 1 \\ 
  123 & 17 & J04394488+2601527 & ITG15 & 69.937 & 26.03 & M5 & 0.178 & II & 1 \\ 
  124 & 17 & J04394748+2601407 & CFHT4 & 69.948 & 26.03 & M7 & 0.057 & II & 1 \\ 
  125 & 17 & - & IRAS04368+2557 & 69.974 & 26.05 & K5c & 1.121 & 0 & 1 \\ 
  126 & 17 & J04400174+2556292 & - & 70.007 & 25.94 & M5.5 & 0.137 & III & 2 \\ 
  127 & 17 & J04400800+2605253 & IRAS04370+2559 & 70.033 & 26.09 & M2c & 0.575 & II & 1 \\ 
  128 & 18 & J04404950+2551191 & JH223 & 70.206 & 25.86 & M2 & 0.575 & II & 2 \\ 
  129 & 18 & J04410424+2557561 & Haro6-32 & 70.268 & 25.97 & M5 & 0.178 & III & 1 \\ 
  130 & 18 & J04410826+2556074 & ITG33A & 70.284 & 25.94 & M3 & 0.398 & II & 1 \\ 
  131 & 18 & J04411078+2555116 & ITG34 & 70.295 & 25.92 & M5.5 & 0.137 & II & 1 \\ 
  132 & 18 & J04411267+2546354 & IRAS04381+2540 & 70.303 & 25.78 & M2c & 0.575 & I & 1 \\ 
  133 & 18 & J04413882+2556267 & IRAS04385+2550 & 70.412 & 25.94 & M0 & 0.701 & II & 1 \\ 
  134 & 19 & J04420548+2522562 & LkHa332/G2A+B & 70.523 & 25.38 & M0 & 0.701 & III & 2 \\ 
  135 & 19 & J04420732+2523032 & LkHa332/G1A+B & 70.531 & 25.38 & M1 & 0.633 & III & 2 \\ 
  136 & 19 & J04420777+2523118 & V955TauA+B & 70.532 & 25.39 & K7 & 0.801 & II & 2 \\ 
  137 & 19 & J04422101+2520343 & CIDA7 & 70.588 & 25.34 & M4.75 & 0.201 & II & 1 \\ 
  138 & 19 & J04423769+2515374 & DPTau & 70.657 & 25.26 & M0.5 & 0.667 & II & 2 \\ 
  139 & 20 & J04551098+3021595 & GMAur & 73.796 & 30.37 & K7 & 0.801 & II & 1 \\ 
  140 & 20 & J04552333+3027366 & - & 73.847 & 30.46 & M6.25 & 0.086 & III & 1 \\ 
  141 & 20 & J04553695+3017553 & LkCa19 & 73.904 & 30.30 & K0 & 2.430 & III & 1 \\ 
  142 & 20 & J04554046+3039057 & - & 73.919 & 30.65 & M5.25 & 0.157 & III & 1 \\ 
  143 & 20 & J04554535+3019389 & - & 73.939 & 30.33 & M4.75 & 0.201 & II & 1 \\ 
  144 & 20 & J04554582+3033043 & ABAur & 73.941 & 30.55 & B9 & 3.250 & II & 1 \\ 
  145 & 20 & J04554757+3028077 & - & 73.948 & 30.47 & M4.75 & 0.201 & III & 2 \\ 
  146 & 20 & J04554820+3030160 & XEST26-052 & 73.951 & 30.50 & M4.5 & 0.225 & III & 1 \\ 
  147 & 20 & J04554969+3019400 & - & 73.957 & 30.33 & M6 & 0.096 & II & 2 \\ 
  148 & 20 & J04555605+3036209 & XEST26-062 & 73.984 & 30.61 & M4 & 0.271 & II & 1 \\ 
  149 & 20 & J04555938+3034015 & SUAur & 73.997 & 30.57 & G2 & 2.632 & II & 1 \\ 
  150 & 20 & J04560118+3026348 & XEST26-071 & 74.005 & 30.44 & M3.5 & 0.335 & II & 1 \\ 
  151 & 20 & J04560201+3021037 & HBC427 & 74.008 & 30.35 & K5 & 1.121 & III & 2 \\ 
    \hline
\end{longtable} }
\tablefoot{This table was built from the catalog in Paper I 
Columns 1--3: Star ID in the table, Identification number of the NEST  2MASS Point Source Catalog and common Name. Column 4--5: Ecliptic right ascension and Declination (Epoch J2000). Columns 6--8: Spectral type, (Primary) mass, Class. Columns 9: Total number of stars ($n_{*}$) within 1000\,au.}
\twocolumn
}

\section{DBSCAN algorithm}
\label{AppSec:Dbscan}

In this appendix, we define more formally the concepts and vocabulary of {\tt dbscan} (Density-Based Spatial Clustering Applications with Noise) clustering method, along with complementary comments relevant to our analysis. Some terms are taken from the work of \cite{EsterEtAl1996}. 
 The $n$ objects of the dataset are assumed to be embedded in 
 a $d$-dimensional space ${ \cal R} \subset { \cal R}^d$. In our analysis of stellar clustering, we will deal with a projected Euclidean space of dimension $d= 2$.

\subsection {Basic concepts of DBSCAN}

The $ \epsilon $-neighborhood of an object $i \in { \cal R}^d $ is defined as the subset $\{ \epsilon \}_i$ of ${ \cal R}^d$ such that : 
\begin{equation}
 \begin{aligned}
\epsilon_i = \{j \in { \cal R}^d \mid d(i,j) \le \epsilon \}
       \end{aligned}
 \qquad \text{$\epsilon$-neighborhood of $i$}
,\end{equation}
\noindent where $d(i,j)$ is the distance between objects $i$ and $j$. We now introduce two intermediate definitions. First, two objects $i$ and $j$  of the dataset are said to be $\epsilon$ direct-neighbors, if their distance $d(i,j)$ is less or equal to  $\epsilon$, in other words, if $j$ belongs to the $\epsilon$-neighborhood of $i$ and vice-versa: 
\begin{equation}
 \left.
 \begin{aligned}
i \in \epsilon_j  \\
j \in \epsilon_i 
  \end{aligned}
 \right \}
 \qquad \text{$\epsilon$-direct-neighbors condition}
.\end{equation}
Secondly, two objects $i$ and $j$  of the dataset are said to be indirect $\epsilon$-neighbors, if we can find a chain $\{l_1,\ldots,l_n\}$ of $\epsilon$-direct-neighbors in ${ \cal R}^d$ connecting $i$ and $j$. So the indirect $\epsilon$-neighbors condition reads as: 
\begin{equation}
 \left.
 \begin{aligned}
     \exists \{l_1,\ldots,l_n \} \in {\cal R}^d \text{ such that }  l_1 = i \text{ and } \,  l_n  = j \\
\forall m= (1,\ldots,n-1), \;(l_m, l_{m+1}) \text{ are } \epsilon\text{-direct neighbors.}\\
       \end{aligned}
 \right \}
\\
 \label{Eq:indirnei} 
\end{equation}
Describing our dataset in terms of vertices (nodes) associated with the objects linked together if $d(i,j) \leq \epsilon$, the direct- and indirect $\epsilon$-neighbors conditions reflect the definition of a path within the graph framework.

The $\epsilon$-neighborhood  population $N_i^{\epsilon}$ of a point $i  \in { \cal R}^d$ is then defined as the number of objects in $ { \cal R}^d$ found within the $\epsilon$-neighborhood of $i$ :
\begin{equation}
 N_i^{\epsilon} = Card(\epsilon_i)
 \qquad \text{$\epsilon$-neighborhood  population of $i$, }
\end{equation}
where $Card$ stands for the cardinality.

To define clusters, a minimal local neighborhood population threshold $N_{min}^{\epsilon}$ is introduced to identify objects of the dataset having at least the same $\epsilon$-neighborhood population or higher. An object $i$ is said to be a core cluster object if its  $\epsilon$-neighborhood  is as populated or even more, than the given local neighborhood  threshold: 
\begin{equation}
 \begin{aligned}
N_i^{\epsilon}  \ge N_{min}^{\epsilon}
       \end{aligned}
 \qquad \text{core object condition.}
  \label{Eq:corecond}
\end{equation}
An object $j$ of the dataset is said to be $(\epsilon,\, N_{min})$ directly density-reachable from $i$ if they are $\epsilon$-direct-neighbors and  if $i$ is a core object:
\begin{equation}
 \left.
 \begin{aligned}
j \in \epsilon_i \\
N_i^{\epsilon}  \ge N_{min}^{\epsilon}. \\
              \end{aligned}
 \right \}
 \qquad \text{direct density-reachability property from i.}
 \label{Eq:ddrp}
\end{equation}
Similarly, a point $j$ is said to be $(\epsilon,\, N_{min})$  density-reachable from $i$, if $i$ and $j$ are $\epsilon$-indirect-neighbors through a chain ${l_1,\ldots,l_{n-1}}$ of core objects.  The density-reachability property reads as:
\begin{equation}
 \left.
 \begin{aligned}
     \exists \{l_1,\ldots,l_n \} \in {\cal R}^d \text{ such that } l_0 = i  \text{ and } \,  l_n  = j, \\
\forall m= (1,\ldots,n-1), \;(l_m, l_{m+1}) \text{ are } \epsilon\text{-direct-neighbors,} \\
\forall m=(1, {\text \ldots},n-1) \, N_{l_m}^{\epsilon}  \ge N_{min}^{\epsilon}.\\
       \end{aligned}
 \right \}
 \qquad 
\end{equation}

Once an arbitrary core cluster object - called  the seed - is identified, the density-reachability property allows to retrieve all the objects of the dataset belonging to the same cluster. A cluster is then defined as the set of points that are density-reachable from a seed.   If an object $i$ is density-reachable from a seed $j$ but does not fulfill the core object condition (eq. \ref{Eq:corecond}), then it is said to be a  border cluster object :
\begin{equation}
 \left.
 \begin{aligned}
    \exists   j \in \{\epsilon\}_i \\
N_j^{\epsilon}  \ge N_{min}^{\epsilon}\\
N_i^{\epsilon}  \nleq N_{min}^{\epsilon}\\
       \end{aligned}
 \right \}
 \qquad \text{border point condition for $i$.}\\
   \label{Eq:bordercond}
\end{equation}
Given $\epsilon$ and $N_{min}^{\epsilon}$, the {\tt dbscan} algorithm proceeds in two steps: (1) pick one object from the dataset, check if it is a seed (i.e., satisfying eq. \ref{Eq:corecond}) and if it is not, pick another one until it satisfies the core object condition; and (2) retrieve all the objects in the dataset which are density-reachable from the seed. We note that the  core points set in  the final clusters obtained at the end of the algorithmic scan does not depend on the chosen seed,  since  the core objects of  a given cluster are  density reachable from each other. A cluster $C$ is defined by  the following conditions:
\begin{equation}
 \left.
 \begin{aligned}
\exists i_{seed} \in {\cal R}^d \text{ such that } i_{seed} \, \text {is a core point of cluster C}  \\
\{ C \} = \{ i  \in {\cal R}^d,  \mid   i  \text{ is  }  (\epsilon,\, N_{min}) \text{ density-reachable from  } i_{seed} \}.\\
       \end{aligned}
 \right \}
 \label{Eq:clustdef}
\end{equation}
If an object of the dataset is not ($\epsilon,\, N_{min}^{\epsilon}$)-density-reachable from any core objects then it is marked as an outlier, or noise, with respect to that property. Noise and outliers are then the set of objects that do not belong to any cluster in the dataset. Based on the above definition and within a graph framework, {\em dbscan clusters} are  in fact the maximal connected component sets $\{ C \}$ such that $\forall (i,j) \in  \{ C \} , \, N_i \geq N_{min}\, $ and there is a path between $i$ and $j$. This definition of clusters is therefore connected to the more intuitive idea of density-contour  clusters outlined by Hartigan (1975).

\subsection {Discussion on {\tt dbscan}}

\subsubsection {Complexity} 
In the general case, the complexity of hierarchical clustering techniques is $\mathcal{O}(n^3)$, which makes them too slow for large data sets. However, but for the single and complete linkage clustering, this is optimally reduced to  $\mathcal{O}(n^2)$.  The {\tt dbscan} clustering algorithm visits each point of the database possibly multiple times (e.g., as candidates to different clusters).  Without the use of an accelerating index structure, the run time complexity is $\mathcal{O}(n^2)$. This can be reduced to $\mathcal{O}(n\log n)$ with a spatial indexing using a R-tree.

\subsubsection {Minimal size of a cluster}

There are two free parameters in the {\tt dbscan} algorithm: $\epsilon$, which defines the size of the neighborhood, or "window", that is investigated around each object, and $N_{min}^{\epsilon}$, the minimal local neighborhood  population threshold, which eventually corresponds to a local density threshold since the population number is defined within a $\epsilon$-neighborhood. 

Since, $N_{min}^{\epsilon}$ is the minimum number of objects that must exist in the $\epsilon$-neighborhood to obtain a core point,  it is thus also the minimum size of a  cluster \citep{EsterEtAl1996}. At least, this minimum cluster is composed by one core point and  $(N_{min}^{\epsilon}-1)$ border points. According to the definition of a  border point (eq. \ref{Eq:bordercond}),  it may happen that a point is  a border point for several different clusters. Then the labeling of this  ambiguous point  to a specific cluster depends arbitrarily on the order to which the points are processed during the algorithmic scan. By convention, it is automatically labeled to the first cluster it meets as border point. The algorithm is one of "exclusive" type, as opposed to non-exclusive clustering algorithm for which  one point may belong to multiple clusters. If this  point appears to be at a later stage also a border point for another cluster and if it is one of the possible smallest cluster, composed then  a priori by $N_{min}^{\epsilon}-1$ border points, this cluster would have consequently in fact less members than  $N_{min}^{\epsilon}$ local population threshold. We note, however, that this fact does not introduce inconsistency with the definition of the density-reachable based cluster (\ref{Eq:clustdef}) although it  highlights the order-dependence of the algorithm. It also outlines two different roles of border points that may vary from being a real frontier that distinguishes a cluster from noise or other clusters from being a   bridge between clusters. 

\subsubsection {Minimum distance between two clusters}

The two parameters $N_{min}^{\epsilon}$ and the $\epsilon$-neighborhood length play a key role in identifying clusters. If no seed point is found, the whole dataset is considered as noise.  Conversely, if a first seed core point is found within the dataset and if all the the points are density-reachable from the seed, we obtain a single cluster containing all objects. The algorithmic  process will give rise to a second cluster when only a subset of all points are associated with a core point and a second core point that is not density reachable from the first seed is found in the set. It is then worthwhile to investigate the two conditions at which the points are not density reachable. The first one states that a point is never  density reachable from a seed point if it is not its $ \epsilon $-indirect neighbor. The second point outlines the fact that a point is never density reachable from a border point, even if this point is in the $\epsilon$-neighborhood of the border point.

So there are two basic situations that stop the ongoing growth of a cluster. One is based on the $\epsilon$-neighborhood length threshold since the algorithm excludes as candidate members all the points that are not $\epsilon$-indirect neighbors of the seed (first condition above). The second is based on $N_{min}^{\epsilon}$ parameter which is the parameter  that is indirectly used in order to decide whether a cluster should be increased at a certain stage of the algorithm. This parameter describes in fact the minimum number of reachable points that a given point must have in its  $\epsilon$-neighborhood  not to be a border point. A significant local drop of $\epsilon$-neighborhood density within the dataset therefore stops the growth process (second condition quoted above). In other words, the cluster growth stops when no more path is found from a core point, when the edges of a cluster are surrounded by less dense regions.
Either way, the first conditions implies that two clusters are separated by a distance of at least $\epsilon$ to discriminate them from being one single cluster. However, this threshold value is only true when the distance between clusters is estimated from core points. This separability distance may indeed  be shorter when  border points are implied in this inter-distance evaluation (second condition). When this occurs, the two clusters are marginally separated due to a localized drop in density. Choosing a shorter $\epsilon$-neighborhood length to scan the dataset would lead to identifying these border points as noise and would separate more clearly the two clusters.

\subsubsection {Range of the number of star components}
\label{Apend_subsub:Range of the number of star components}

Given the $\epsilon$-neighborhood length parameter, the $N$ objects of the dataset are partitioned into  a set $\{O\}$ grouping data that are noise or outliers with $0\le N_O=Card(\{O\}) \le N$ and  $N_C$ number of clusters. When no seed cluster is found, the condition eq. \ref{Eq:corecond} is never fulfilled, no cluster is detected and $N_O=N$ and  $N_C=0$. Excluding the pathological cases implying "ambiguous points", the theoretical maximum value of the number of the clusters $N_C$ that can be found to partition the dataset is obtained when the neighborhood of each core point is composed by ($N_{min}^{\epsilon}-1$) border points. The total population of neighborhood is then exactly equals to the minimal value $N_{min}^{\epsilon}$. Then, the maximal range of the number of clusters is given by  $ 0\le N_C \le E(N/N_{min}^{\epsilon})] $ where $E$ is the floor function.

\section{Discussion on clustering algorithms} 
\label{AppSec:ChoiceClustAlgo}

This appendix is intended to review the different types of clustering algorithms in order to provide the context for our choice of the one level {\tt dbscan} algorithm as the optimal procedure to detect local overdense spatial structures. 

Clustering is the process of examining a collection of objects embedded in a $d$-dimensional space to group   {\em  closest} objects together automatically  into {\em natural clusters}, while distinguishing groups from each other. Although this task may sometimes be easily done by visual inspection in a 2-dimensional space, it is not straightforward to formalize the process on general grounds. The fundamental challenge is to conceptually and quantitatively define the meaning of {\em close}, hence to define from heterogeneous variables a proximity (or distance) function between objects and to define a meaningful cluster membership criteria to assign (or not) objects in clusters. This is why many clustering algorithms have been developed over the last five decades \citep[for general reviews, see][]{KaufmanRousseeuw1990,TanEtAl2005,XuEtAl2005,Jain2010,EverittEtAl2011,MurtaghContreras2012} and the choice for a given study must be guided by the nature of the available datas and scientific objectives.

\subsection{Hierarchical clustering} \label{AppSubSec:HieraClustAlgo}

Standard hierarchical clustering methods group data over a variety of scales by creating a cluster binary tree, or dendrogram. The tree is not a single set of clusters, but rather a multilevel hierarchy, where clusters at one level are joined as clusters at the next level. It is left to the user to decide and quantitatively define the scale of clustering (i.e., the cut-off level) that is ``most appropriate.'' Of course, this definition is a matter of debate since there is no general law, so at best the criteria used are heuristic, and ultimately will depend on the clustering purpose of the user. 

A traditional class of hierarchical clustering algorithms are the so-called agglomerative, or bottom-up, algorithms. These non-parametric algorithms start from initial set of clusters (each composed of a single object) and iteratively build nested families of larger clusters through successive mergers of cluster pairs, using a heuristic proximity function criteria, until one single cluster containing all objects in the distribution is obtained. 

The other traditional class of clustering algorithms are the non-parametric divisive hierarchical clustering algorithms, which essentially are the reverse of the agglomerative algorithms. That is to say that such algorithms start from the entire set and, step by step, divide the set in subsets until single objects are obtained. This approach is fundamentally related to the previous one, only being divisive instead of agglomerative. However, since the divisive approach is somewhat more complicated to implement \citep[for a review on spectral partitioning, see][]{Filippone2008}, here we focus on the agglomerative algorithms.

In agglomerative algorithms a similarity (i.e., proximity) or dissimilarity (i.e., distance) function is defined heuristically at each merging step as described by the linkage method being employed, which is selected by the user prior to implementing the algorithm. In this approach, six most widely employed linkage methods are generally used, depending on the distance criteria chosen to proceed to the merging of clusters. 
Firstly the "single Link" method has a merging criteria based on the distance between the closest single element pair of two clusters in the set. Secondly, the "complete link" method is based on the distance between the furthest element pair of two clusters in the set. Thirdly, the "average link" method is based on the distance between the each element pair of two clusters in the set. The other two methods,   "centroid" and "median" links are respectively based on  the distance between the centroids of two clusters and the median distance between the each element pair of two clusters in the set. Although similar to the average link method, the median link method is less sensitive to outliers. Finally, the sixth method is the  "Wards" link based on the value of the sum of squared deviations between the projected centroid of two clusters in the set. 

In all these bottom-upx< methods, the cluster pair in the set with the smallest distance are then merged. The strength of these non-parametric methods is that no assumptions are made about the structure of data. However, they suffer from a significant drawback: since each step of the agglomerative process is built upon the previous one, and no backtracking is permitted, there is a loss of data such that the nested hierarchical structure is built-in and not the result of an open process free of reconfiguring the structure of data at each step. In other words, the hierarchical clustering algorithm is sensitive to possibly erroneous previous cluster merging, since, once assigned, objects' cluster memberships are not permitted to change (i.e., the assignments at each step are permanent). 

Another significant drawback of these agglomerative methods is their lack of robustness, especially when the distribution contains noise and outliers. Generally speaking, this leads spurious additional clusters being identified or the blurring of distinct clusters. It also tends to produce clusters of a particular shape. For example, in the single linkage method, the merging of two clusters relies only on a local merging criteria based on their most proximate members. Thus, only the closest parts of the two clusters are considered whereas the overall structure of the clusters and their more distant parts are not taken into account. Since clusters are merged on the criteria of the closest element pair of two clusters in a set, it may happen that in a noisy distribution elements are successively merged in such a way that two obvious distinct clusters are merged together (due to few sparse singletons being close to each other), even though many of the elements in each cluster are very distant from each other. This tendency to combine elements linked by a series of close intermediate elements gives rise to long chains (this is known as the chaining effect). On the other hand, in the complete linkage methods, the merging of two clusters at one step is based on the proximity of their furthest members. Thus, it tends to choosing at each step, the merger of the pair of clusters that gives rise to a final cluster that has the smallest diameter. And since the diameter of the merged clusters is minimized at each iteration, the method has a tendency to give rise to compact "globular" (dense and circular or spherical) clusters. This criterion for merging in complete link methods is non-local in the sense that the entire distribution of members in a cluster can influence the choice of the merging. A single element located far away from the majority of the members of one cluster increases significantly the final diameter of two merged clusters which, in turn, may lead to a major change in the final clustering. Thus, the complete-link methods are rather sensitive to outliers, as the single-link methods are sensitive to the noise. 

Closely related to the agglomerative linkage methods, the Minimum Spanning Tree (MST) method is an alternate algorithm to build clusters from a given length threshold. When considering a set of spatial points as a complete graph, in which the points are vertices and edges are the euclidean distance between vertices, the MST is the subgraph in which any two vertices are connected by exactly one simple path (i.e., a connected graph without cycles, by definition a tree) such that the total distance length of its edges is minimal. In this framework, clusters are defined as the remaining connected graphs when deleting the largest lengths of MST above a heuristically chosen threshold \citep[inconsistent edges,][]{Zahn1971}, or the critical length  \citep{GutermuthEtAl2009}, such that clusters are then the remaining connected subgraphs (i.e.,  there is a path between any two points of the subgraph).  The same clusters are obtained from the linkage methods agglomerative algorithms employ by cutting the dendrogram at the same critical length, so that each connected component forms a cluster. Indeed, it is well established that the same information required to build the MST of a set of points is contained within the dendrogram generated by the linkage methods (see e.g., \cite{GowerRoss1969}.  Therefore, the same drawbacks (high sensitivity to noise and outliers that blur cluster structure) affect both the MST and linkage methods.

\subsection{Partitioning clustering} \label{AppSubSec:ParClustAlgo}

Iterative relocation algorithms use an iterative control strategy to optimize the partition of distribution into clusters.  The number of clusters is usually an input parameter for these algorithms, in other words some a priori domain assumption knowledge is required \citep[for a review, see][]{Mirkin2005}. When the number of clusters is given, there are two statistical approaches in order to partition the data in these clusters:
  
Non-parametric iterative relation algorithms (K-means, K-median) are based on the assumption that clusters correspond to different modes of the probability density, in other words, they are associated with the values that appear more often in a data set. Once an initial partition is given, each cluster is associated with its representative, e.g., its barycenter, centroid, mean point, or by one of the objects of the cluster located near its center. The goal of this class of algorithm is to assign each object to its closest representative and iteratively process to minimize the objective function, as for example the distance of elements to the their closest representatives summed over the over whole set; the iterative process stops when convergence on representatives is reached. For example, the K-means algorithm \citep{KanungoEtAl2002}, a variance-based function, can be shown to minimize within-cluster distance while maximizing between-cluster distance. It should be noted that the assignment of each object to a cluster implies that the induced partition is equivalent to a Voronoi diagram \citep{InabaEtAl1994,ImaiInaba1995}. Thus, the shape of clusters found by a partitioning algorithm is always convex which is a restrictive bias of the method. 

On the other hand, parametric iterative relation algorithms (mixture models) are based on the assumption that each cluster is represented by a set of parametric probability density coming from a same family. For example, either a traditional Gaussian distribution, or an "isothermal ellipsoid" \citep{KuhnEtAl2014}, would be employed as a component mixture model method. The "parametric" term refers to the choice of function representing a cluster. Unlike the previous methods, the number of the clusters is required as a given input prior the clustering.


For this class of iterative relocation algorithms, the goal is to estimate (1) the number (and type of  the geometrical distribution in the case of parametric methods) of clusters (structure identification), and (2) the parameters of the distributions (weight, variance and mean for each probability distribution associated with a cluster). Broadly speaking, partitioning cluster analysis algorithms are then based on  two distinct phases: first is a model fitting phase, whereby the number and the geometrical type of component are a priori chosen, and second is a model validation phase, whereby the set of data are assessed to each cluster according to some cluster validity criterion and to one "optimal" partition hypothesis selected through a maximization technique of an objective function. In most cases, the expectation-Maximization (EM) technique of the (log-)likelihood function is used. The EM technique is able to propose the best local solution in order to smoothly distribute data in each cluster and to constrain the parameters of each probability distribution associated with a cluster. For example, the K-means approach is a special case of EM clustering applied to a mixture of Gaussians when all variances are equal \citep{GuptaChen2011}. This is why the K-means method leads to globular shaped clusters.

Iterative relocation methods have several drawbacks: the number of clusters or components have to be set prior to implementation. They are sensitive to first initialization of the partition. A  guess of the number of clusters can possibly be inferred based on the Akaike Information Criterion (AIC) such as used in the work of \cite{KuhnEtAl2014}. The AIC is a probabilistic indicator that somehow evaluates the gain of introducing some supplementary cluster, and then all the associated free parameters, to improve the model clustering fit evaluated through the global maximum (log-)likelihood function at the expense of model parsimony. Although it is a quantitative criteria, the derived clusters are biased toward convex (i.e., globular) shaped clusters and most importantly, they do not identify arbitrarily sized clusters since they are biased to equipartition. The notion of noisy data is also not considered, in other words, it is a termination requirement that every single object has to be assigned to a cluster.  

More sophisticated parametric approaches have been developed to deal with the structure identification based on the optimization of the Bayesian Information Criteria that allows the choice of the "best" model based on direct models comparison, models that are computed simultaneously, each combining different weight of clustering methods (hierarchical and EM partitioning) and various number of components \citep{FraleyRaftery1998,FraleyRaftery2002,FraleyRaftery2007}. But even these models do not address the issue of the possible skewness of data distribution in which a single, skewed (or elongated) distribution is described by multiple normal distributions. Moreover, even combining hierarchical and partitional clustering, since there is no explicit notion of noise, the "best" model will be chosen based on complete partition of data even if data are sub-structured, which may be in fact a serious obstacle in determining locally small but significant over-densities.  

For these reasons, the iterative relocation partitioning clustering algorithms do not appear adapted to the objective of our study, especially in view of the fact that these algorithms result in a single complete partition of stars into a unique set of disjoint clusters, which may not suit our perception of young star clusters topology that is most of the time complex, highly inhomogeneous and sub-structured. In summary, given a point-like distribution,  the iterative relocation algorithms imposed on by the parametric cluster model, and by the implicit complete partitioning of,  prevent the identification of non-globular, irregularly shaped and local small-scale substructures. Hence, partitioning algorithms are ruled out here, and the goal of identifying substructures as local over-densities drives us to the third class of partitioning clustering methods, the density based techniques.

\subsection{Density based clustering} 

Density clustering algorithms, such as {\tt dbscan} \citep{EsterEtAl1996}, partition a distribution into clusters based on a density criterion. Essentially, they are the formalization of the assumption that clusters are  high-density regions surrounded by low-density regions \citep{Hartigan1975}. These algorithms borrow (1) some aspects of the non-parametric and parametric iterative partition cluster methods, and  (2) some aspects of MST (single linkage) hierarchical clustering methods. For the former, density clustering algorithms are a non-parametric method, as there is no a priori given function associated with cluster postulating a specific structure for discrete distributions. However, the notion of density probability is not completely ignored since there is an estimation of the local density from a kernel density estimator associated with a given smoothing bandwidth that reduces the complexity. The non-parametric estimation of probability density functions is based on the concept that the value of a density function at a continuity point can be estimated using the sample observations that fall within a small region around that point, known as the Parzen kernel class of density estimates. For a comprehensive review of kernel non-parametric density estimation and the crucial issue of the smoothing parameter (bandwidth) choice see for example  \cite{Silverman1986} and \cite{Scott1992}. In the case of the dbscan algorithm, a given circular and uniform kernel acts as a given smoothing bandwidth ($\epsilon$-neighborhood for dbscan) kernel.

For {\tt dbscan}, clusters are also defined as the maximal connected subgraphs remaining at a cutting scale length threshold, but these subgraphs are composed by selected instances (thus allowing an incomplete partitioning). A significant advantage of using such density based clustering algorithms is that the cluster membership criterion is based on a simple density requirement that allows noisy data to be filtered out, whereas this cannot be done using hierarchical and iterative relocation based partition based algorithms. The {\tt dbscan} algorithmic implementation appears particularly well suited to our natural perception of star clusters and the study of their structure.
          
\end{document}